\documentclass[prx,aps,epsf,showpacs,twocolumn,longbibliography]{revtex4-1}
\usepackage{times}
\usepackage{graphicx}
\usepackage{float}
\usepackage{latexsym,amsmath,amssymb,bm,euscript}
\usepackage{color}
\usepackage[justification=raggedright]{caption}
\usepackage{subcaption}
\usepackage{epstopdf}
\usepackage[colorlinks=true,linkcolor=blue,citecolor=blue,urlcolor=blue]{hyperref}
\usepackage{hyperref}

\usepackage{physics}

\newcommand{\br}{{\bf r}}

\newcommand{\bk}{{\bf k}}

\DeclareMathAlphabet{\mathpzc}{OT1}{pzc}{m}{it}

\newcommand{\be}{\begin{eqnarray}}
\newcommand{\ee}{\end{eqnarray}}
\newcommand{\bw}{\begin{widetext}}
\newcommand{\ew}{\end{widetext}}

\usepackage{custom_commands}

\begin{document}

\title{Quantum Anomalous Hall Insulator Stabilized By Competing Interactions}
\author{Shouvik Sur$^1$}
\author{Shou-Shu Gong$^{2,3}$}
\email{shoushu.gong@buaa.edu.cn}
\author{Kun Yang$^1$} 
\author{Oskar Vafek$^1$}
\email{vafek@magnet.fsu.edu}
\affiliation{
$^1$National High Magnetic Field Laboratory and Department of Physics, Florida State University, Tallahassee, Florida 32306, USA\\
$^2$Department of Physics, Beihang University, Beijing 100191, China\\
$^3$National High Magnetic Field Laboratory, Florida State University, Tallahassee, Florida 32310, USA
}
\date{\today}
\begin{abstract} 
We study the quantum phases driven by interaction in a semimetal with a quadratic band touching at the Fermi level.
By combining the density matrix renormalization group (DMRG), analytical power expanded Gibbs potential method, and the weak coupling renormalization group, we study a spinless fermion system on a checkerboard lattice at half-filling, which has a quadratic band touching in the absence of interaction.
In the presence of strong nearest-neighbor ($V_1$) and next-nearest-neighbor ($V_2$) interactions, we identify a site nematic insulator phase, a stripe insulator phase, and a phase separation region, in agreement with the phase diagram obtained analytically in the strong coupling limit (i.e. in the absence of fermion hopping).
In the intermediate interaction regime, we establish a quantum anomalous Hall phase in the DMRG as evidenced by the spontaneous time-reversal symmetry breaking and the appearance of a quantized Chern number $C = 1$.
For weak interaction, we utilize the power expanded Gibbs potential method that treats $V_1$ and $V_2$ on equal footing, as well as the weak coupling renormalization group.
Our analytical results reveal that not only the repulsive $V_1$ interaction, but also the $V_2$ interaction (both repulsive and attractive), can drive the quantum anomalous Hall phase.
We also determine the phase boundary in the $V_1$-$V_2$ plane that separates the semimetal from the quantum anomalous Hall state.
Finally, we show that the nematic semimetal, which was proposed for $|V_2| \gg V_1$ at weak coupling in a previous study, is absent, and the quantum anomalous Hall state is the only weak coupling instability of the spinless quadratic band touching semimetal.
\end{abstract}

\maketitle

\section{Introduction}
The integer quantum Hall state is a paradigmatic example of a topologically non-trivial phase of matter that is realized in the absence of time-reversal symmetry~\cite{prange1990}.
In conventional integer quantum Hall systems time-reversal symmetry is explicitly broken by an externally applied magnetic field, and its topological origin is revealed by the quantized Hall conductivity which is a physical consequence of the non-trivial Chern number that characterizes integer quantum Hall states~\cite{thouless1982}. 
In the integer quantum Hall state the single-particle spectrum is gapped in the bulk, while it remains gapless at the edges due to topological protection.
An externally applied magnetic field, however, is not necessary for the existence of an integer quantum Hall state as demonstrated theoretically by Haldane~\cite{haldane1988}, and simulated in ultracold atom experiments~\cite{jotzu2014}.
This new type of integer quantum Hall state realized in the absence of a magnetic field is called a quantum anomalous Hall (QAH) state. 
In the QAH phase the Chern number is non-trivial and leads to topologically protected gapless edge states.

QAH states resulting from breaking time-reversal symmetry through magnetic doping~\cite{yu2010} or intrinsic ferromagnetism~\cite{liang2013} has been discussed extensively, and realized experimentally \cite{chang2013, checkelsky2014, chang2015}.
An alternative route for realizing a QAH state is through interaction driven spontaneous time-reversal symmetry breaking. 
Such QAH orderings have been argued to exist in two dimensional semimetals with vanishing~\cite{raghu2008}, as well as finite~\cite{sun2009, nandkishore2010,liang2017} density of states at the Fermi level.
While some mean-field based analyses propose the presence of a QAH state at finite interaction strength in Dirac semimetals~\cite{raghu2008,weeks2010,grushin2013,djuric2014}, other analytical and numerical studies find charge ordered  phases instead  \cite{garcia2013,jia2013,daghofer2014,guo2014,motruk2015,capponi2015,scherer2015}. 
Although the QAH phase appears to be absent for linearly dispersing  fermions on the honeycomb lattice, other routes for stabilizing a QAH state have been explored~\cite{zhang2011,ruegg2011,pereg2012,kurita2016,kitamura2015,wang2015,venderbos2016,venderbos2016-2}.
One such route utilizes the finite density of states at the  Fermi level in two-dimensional semimetals with a quadratic band touching  point (QBT)~\cite{sun2009, nandkishore2010}. 
Due to a finite density of states, nearest-neighbor repulsive interaction, $V_1$, is marginally relevant and can drive 
weak coupling instabilities in the semimetal \cite{chong2008,sun2008,nandkishore2010}.
The instability is accompanied by a spontaneous breaking of one of the symmetries that protect the QBT. 
Although the runaway flow can potentially lead to distinct symmetry broken states, energetics imply that  the QAH state is the dominant instability in spinless fermion system~\cite{sun2009,wen2010,tsai2015}.
We note that for attractive interactions, due to an absence of a Fermi surface, the pairing channel mixes with various particle-hole scattering channels which  suppresses superconductivity~\cite{murray2014, vafek2010}.

Notwithstanding the promise of the analytic results, they cannot rigorously establish the presence of the QAH state because on the one hand a runaway renormalization group (RG) flow leads to a loss of analytic  control over RG based predictions, and on the other hand mean-field based results are reliable only in the presence of weak quantum fluctuations which are excluded a priori from such analysis.
Therefore, numerical analyses become essential for unambiguously  establishing the presence of the QAH phase.
Owing to its origin in a marginally relevant operator, the putative QAH gap has the BCS form \cite{sun2009}, and grows exponentially slowly such that at weak coupling the gap is usually too small for numerical detection on finite-size systems.
At strong interaction, however, classical charge ordered  states are stabilized \cite{nishimoto2010,pollmann2014}.
This leaves a small window along the interaction axis for a numerical detection of the QAH gap.
While exact diagonalization calculations find evidence supporting the presence of a QAH phase in the  checkerboard lattice model \cite{wu2016}, fully establishing the nature of the phase within this window remains a challenge due to limitations on the system-size. 
Thus the identification of the QAH phase driven by $V_1$ interaction remains an open question.

Recently, by considering not only $V_1$ but also further-neighbor repulsive interactions such as second- and third-neighbor interactions, numerical calculations have established a QAH phase in various lattice models of spinless fermions~\cite{zhu2016,gongaps2017,chen2018}. 
The QBT realized in the kagome-lattice and decorated-honeycomb-lattice models, however, host a flat valence band which leads to a lack of particle-hole symmetry, generally requires fine tuning to maintain the flatness, and non-generically enhances the effects of interactions. 
Moreover, due to the correlation length exceeding the system size near a continuous phase transition, 
numerical simulations suffer from finite-size effects 
at weaker couplings.
Thus the fate of systems with further-neighbor interactions is unclear closer to the non-interacting point on the phase diagram.
In particular, it is not obvious that the QAH state predicted from weak-coupling RG analysis of the $V_1$ interaction is identical to the one obtained numerically at intermediate-coupling in the presence of further-neighbor interactions.
Further there is always a possibility for some other symmetry broken state to exist at intermediate couplings in a multidimensional coupling space.
Since in models of spinless fermions further-neighbor interactions result in derivative coupling in the low-energy effective theory, an asymptotic analysis is difficult in the presence of such operators which introduce sensitivity to lattice physics.
Moreover, a mean-field description is hindered by a lack of direct decomposition of the further-neighbor interactions into local order parameters defined on the nearest-neighbor sites.

\begin{figure}[!t]
\centering
\begin{subfigure}[b]{0.99\columnwidth}
\includegraphics[width=0.6\columnwidth]{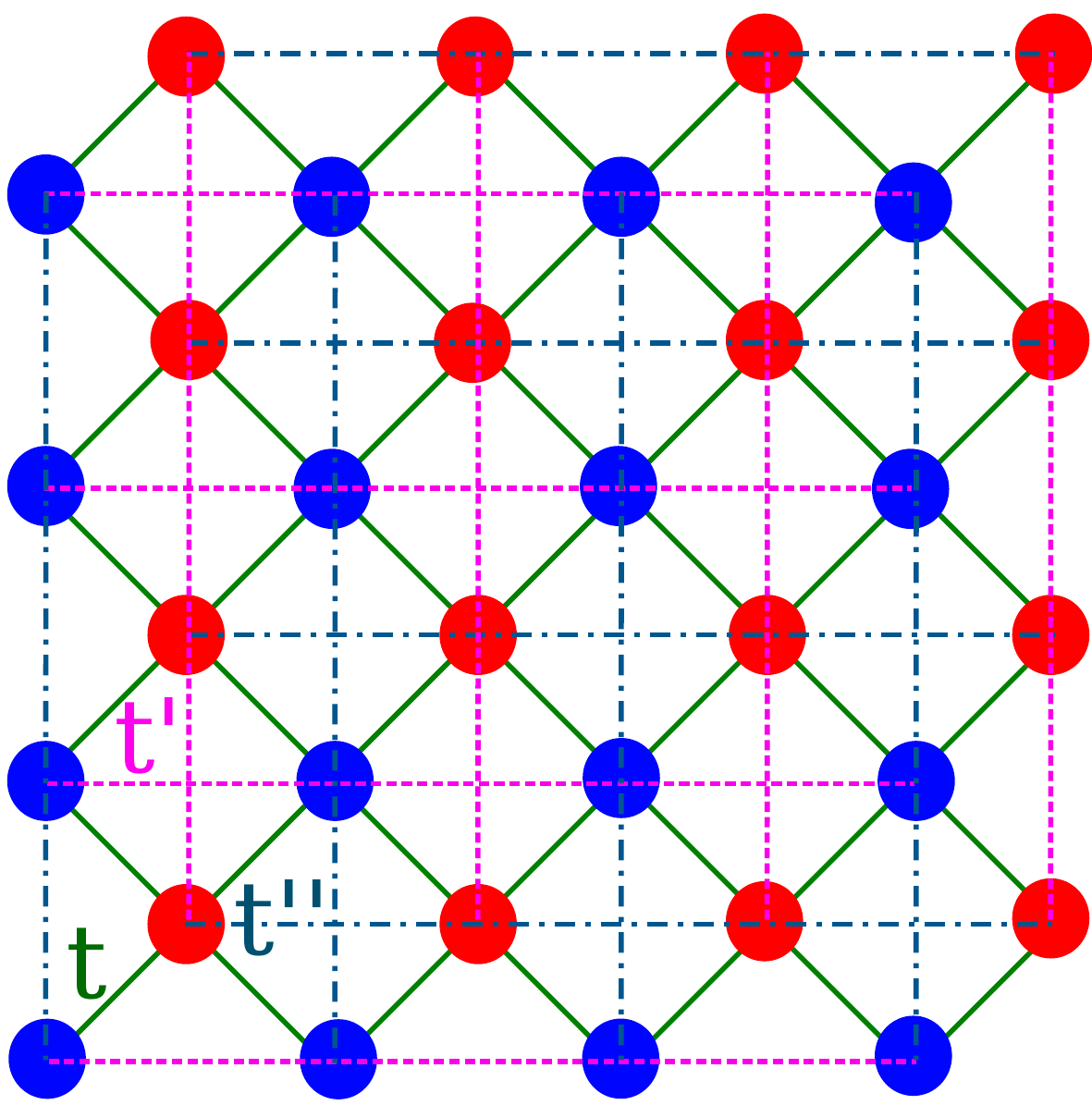}
\caption{}
\end{subfigure}
\hfill
\begin{subfigure}[b]{0.9\columnwidth}
\includegraphics[width=0.9\columnwidth]{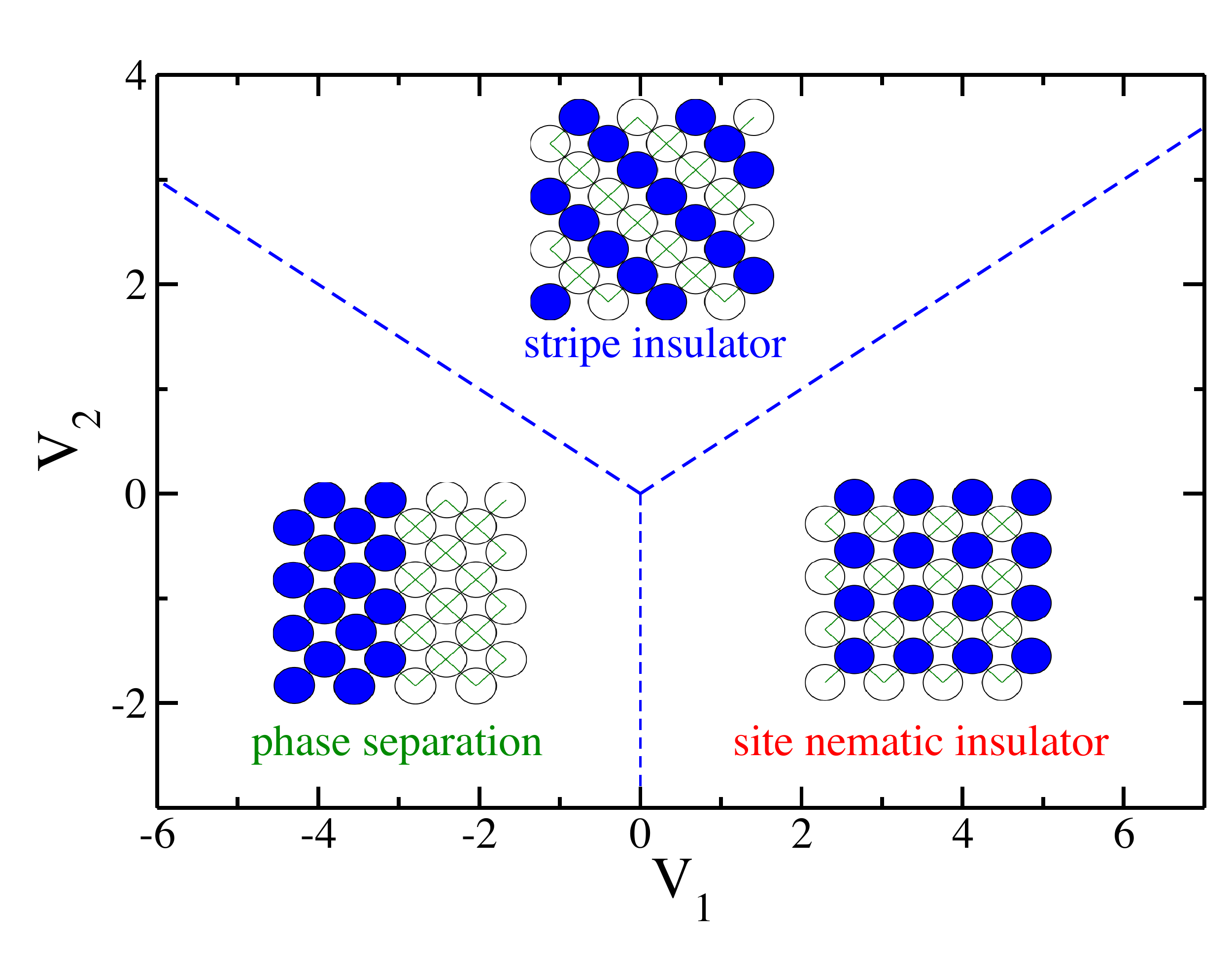}
\caption{}
\end{subfigure}
\caption{Model Hamiltonian and ``classical" phase diagram of the spinless fermion model, Eq.~\eqref{eq:ham}, on the checkerboard lattice. (a) Schematic figure of the model on the $L_y = 4, L_x = 4$ checkerboard lattice. The blue and red dots denote the two sublattices. The (green) solid lines are the nearest-neighbor hopping $t$ between the sites of different sublattices. The (red) dashed lines and the (blue) dashed-dot lines represent the next-nearest-neighbor hopplings $t' > 0$ and $t'' < 0$ between the sites in the same sublattice. While the blue sublattice has $t'$ ($t''$) along the $x$ ($y$) direction, the red sublattice has the opposite choice. We consider the nearest-neighbor ($V_1$) and the next-nearest-neighbor ($V_2$) density-density interactions. (b) Classical phase diagram of the checkerboard-lattice model at half filling. Here, ``classical'' implies an absence of the hopping terms. 
With changing interactions, the model has three insulating phases including the site nematic 
insulator, the stripe insulator, and the phase separation, whose 
schematic figures are shown in the inset with the solid (hollow) circles denoting the 
fully occupied (empty) sites. The dashed lines denote the phase boundaries between these 
insulating phases.
They are  obtained by comparing the energy of each state, viz.  $E_{\rm nematic} = V_2$, $E_{\rm stripe} = V_1/2$, $E_{\rm phase-separation} = V_1 + V_2$.
}
\label{fig:model}
\end{figure}

In this paper we will address the above issues by a combination of analytical and numerical methods.
For concreteness we consider an interacting spinless fermion model on the  checkerboard lattice which is governed by the Hamiltonian,
\begin{equation}
H = -\sum_{ij} (t_{ij} c^{\dagger}_{i}c_j + h.c.) + V_1 \sum_{\langle ij\rangle} n_i n_j 
+ V_2 \sum_{\langle\langle ij\rangle\rangle} n_i n_j,
\label{eq:ham}
\end{equation}
where $t$ is the nearest-neighbor hopping, $t'$ and $t''$ are the next-nearest-neighbor hoppings along two lattice spacing directions [see Fig.~\ref{fig:model}(a)], and $V_1$ ($V_2$) is the nearest-neighbor (next-nearest-neighbor) interaction.
We use $t$ to set the energy scale, and fix $t = 1$.
The Hamiltonian is invariant under discrete translation, time-reversal, and fourfold ($C_4$) rotation.
By setting $t' = -t''$ it acquires a particle hole symmetry as well.
For convenience we choose $t' = 0.5$.
Without interaction a QBT is realized at half-filling. 
The nearest-neighbor interaction, $V_1$, directly leads to a marginal operator in the low energy effective theory, and destabilizes the semimetal when it is repulsive~\cite{sun2009,murray2014}.
At strong coupling, however, $V_1$ leads to a localized state -- the site nematic insulator -- which  spontaneously breaks the $C_4$  symmetry.
The presence of a distinct symmetry broken state at stronger coupling complicates the numerical determination of the QAH state in finite-size systems.
Since a strong repulsive next-nearest-neighbor interaction, $V_2$, stabilizes a   different localized state -- the stripe insulator -- as shown in Fig.~\ref{fig:model}(b), in the presence of both $V_1>0$ and $V_2>0$, quantum fluctuations are enhanced through a mutual frustration of the respective localized states.
This may broaden the window for the realization of a quantum liquid state.
Indeed with a large-scale density matrix renormalization group (DMRG) calculation we report an unambiguous detection of the QAH state on the checkerboard-lattice model as shown in \fig{fig:phase}.
We provide details of the numerical calculation and results in Section \ref{sec:dmrg}.
This is  one of the main results of the paper.

Although a QAH state is detected around $V_1 \sim V_2^2 \sim 4$, all the symmetry broken states within the central triangular region of \fig{fig:phase} may not be QAH since the non-interacting QBT is susceptible towards nematic semimetallic states that break the $C_4$ rotational symmetry down to $C_2$ and compete with the QAH state~\cite{sun2009}.
In order to compare the symmetry broken states obtained in the weak-coupling region of the phase diagram to the numerically determined QAH phase, in Section \ref{sec:pegp} we introduce an analytical method, \emph{power expanded Gibbs potential} (PEGP) \cite{plefka1982}, that treats $V_1$ and $V_2$ on equal footing. 
By utilizing the PEGP we determine the phase diagram in the neighborhood of the QBT, and identify the phase boundary that separates the QBT semimetal from the QAH state.
In the presence of interaction the susceptibilities of the QAH state and the two nematic states diverge along the runaway flow.
The rates of divergence of susceptibilities, however, are distinct, and the nematic semimetals remain subdominant to the QAH state as shown in Section~\ref{sec:nematic}.
In the same section we provide additional support to the susceptibility analysis with  PEGP and numerical calculations.
Our conclusion differs from  Refs.~\cite{sun2009,fradkinBook} in that we do not find a nematic semimetal state at weak coupling, and the QAH state is the sole instability of the QBT in the presence of further-neighbor interaction.
The combined numerical and analytic results strongly suggest that the QAH phase driven by weak interactions  extends to intermediate  interaction region, and  competing further-neighbor interactions play an important role in stabilizing the QAH state.

\section{Numerical determination of the quantum phase diagram} \label{sec:dmrg}
\begin{figure}[!t]
\includegraphics[width=1\linewidth]{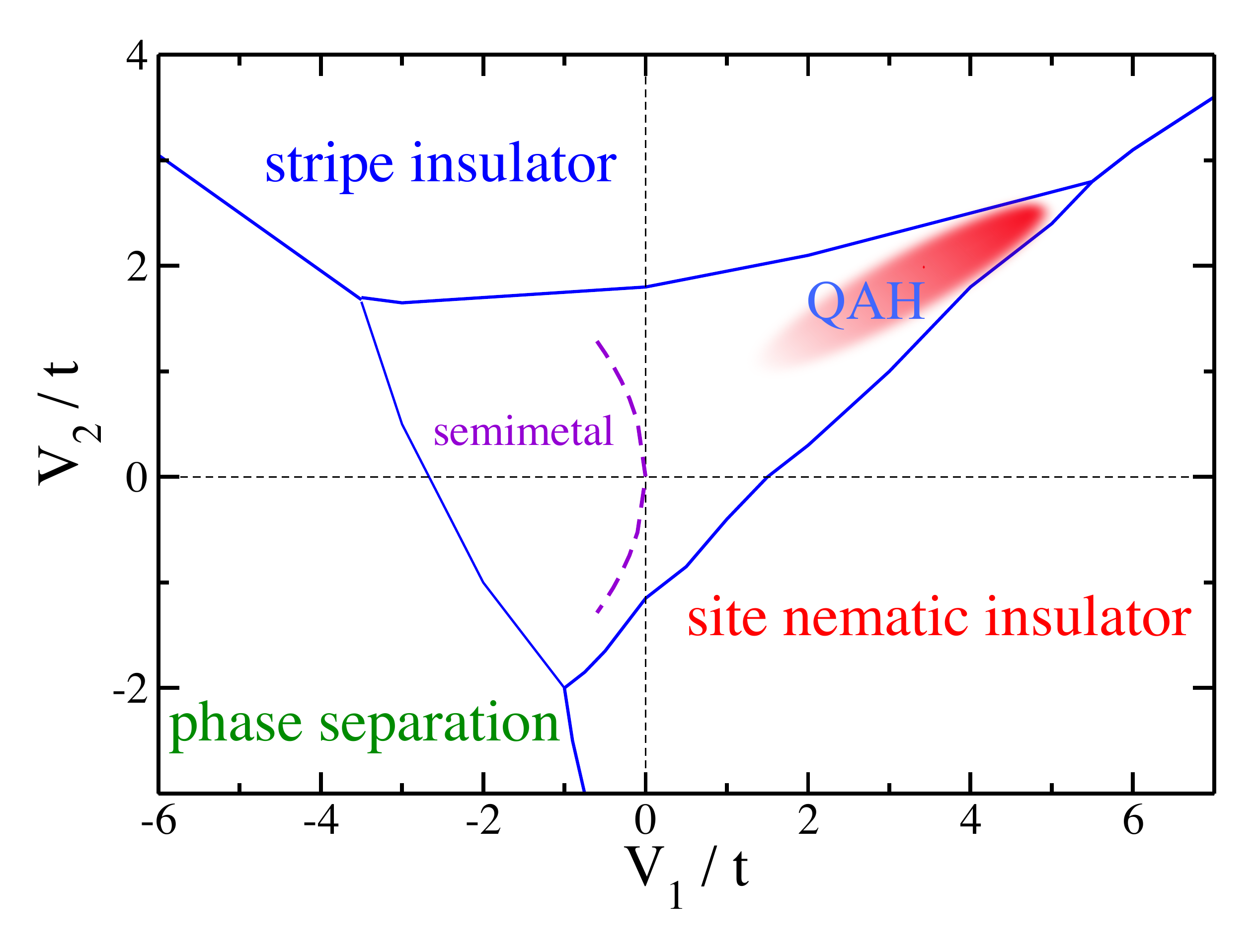}
\caption{Quantum phase diagram of the spinless fermion model, Eq.~\eqref{eq:ham}, on the checkerboard lattice with half filling. In this phase diagram, we set $t'/t = 0.5, t''/t = -0.5$. By tuning  $V_1$ and $V_2$ we identify the ``classical'' insulating phases at strong interaction, which are consistent with the classical phase diagram Fig.~\ref{fig:model}(b). In the central triangular region enclosed by the classical phases we do not find any charge ordered  order. In the shaded region between the site-nematic insulator and the stripe insulator phase, we identify a QAH phase in DMRG calculation as discussed in Section \ref{sec:dmrg}. The QAH state spontaneously breaks time-reversal symmetry and possesses a quantized topological Chern number $C = 1$. Since the system size in the $\hat y$-direction is limited in DMRG calculation, we are unable to distinguish between the QBT semimetal and a weak QAH phase. Using the PEGP method (see Section \ref{sec:pegp}) we find that besides the repulsive $V_1$ interaction, $V_2$ interaction (both repulsive and attractive) can also stabilize a QAH phase. We obtain the dashed line, $V_1 \sim - V_2^2$, separating the QAH phase from the semimetal from the low-energy effective theory. As shown in Section \ref{sec:nematic}, nematic semimetal states that compete with the QAH state remain subdominant and do not appear in the weak coupling region of the phase diagram.
}
\label{fig:phase}
\end{figure}
In this section, we use the unbiased DMRG~\cite{white1992} method to study the model in Eq.~\eqref{eq:ham}.
In DMRG calculations, the numerical accuracy can be controlled by the number of optimal states retained, and the system size can be much larger than that in exact diagonalization calculation which  significantly reduces  finite-size effects.
We consider a cylindrical geometry for the system with periodic boundary conditions along the $y$ direction and open boundary conditions along the $x$ direction.
We illustrate the choice in Fig.~\ref{fig:model}(a) with $L_y$ and $L_x$ denoting the numbers of unit cells along the $y$ and $x$ directions, respectively.
Our system size is up to $L_y = 8$, while $L_x$ is usually taken from $48$ to $64$.
We keep  up to $4000$ optimal states and obtain  very accurate results for $L_y = 4$ and $6$, and convergence to within  truncation errors less than $5 \times 10^{-5}$ for $L_y = 8$.


We determine the quantum phase diagram in Fig.~\ref{fig:phase} in the presence of the hopping terms. 
Our DMRG calculations identify the insulating charge ordered  phases in the strong $(V_1, V_2)$ region, which are separated by the solid-line phase boundaries (for computational details see Appendix~\ref{app:cdw}). 
At large interactions quantum fluctuations due to the hoppings are suppressed, and the quantum phase boundaries approach the ``classical" ones in \fig{fig:model}. 
It is in principle possible to realize a region of coexistence of the QAH and a nematic semimetal in the neighborhood of the non-classical phase boundaries~\cite{sun2009}.
In this work we do not study this possible coexistence region in detail.

Within the central triangular region abutting the charge ordered  phases our DMRG calculations unambiguously identify a QAH phase with spontaneous time-reversal symmetry breaking and quantized Chern number $C = 1$ in the region where $V_1 \sim V_2^2$. 
In the rest of this section  we provide  numerical evidences  for establishing the QAH phase.

\subsection{Spontaneous time-reversal symmetry breaking}
\label{sec:TRS}
\begin{figure}[t]
\includegraphics[width=1.0\linewidth]{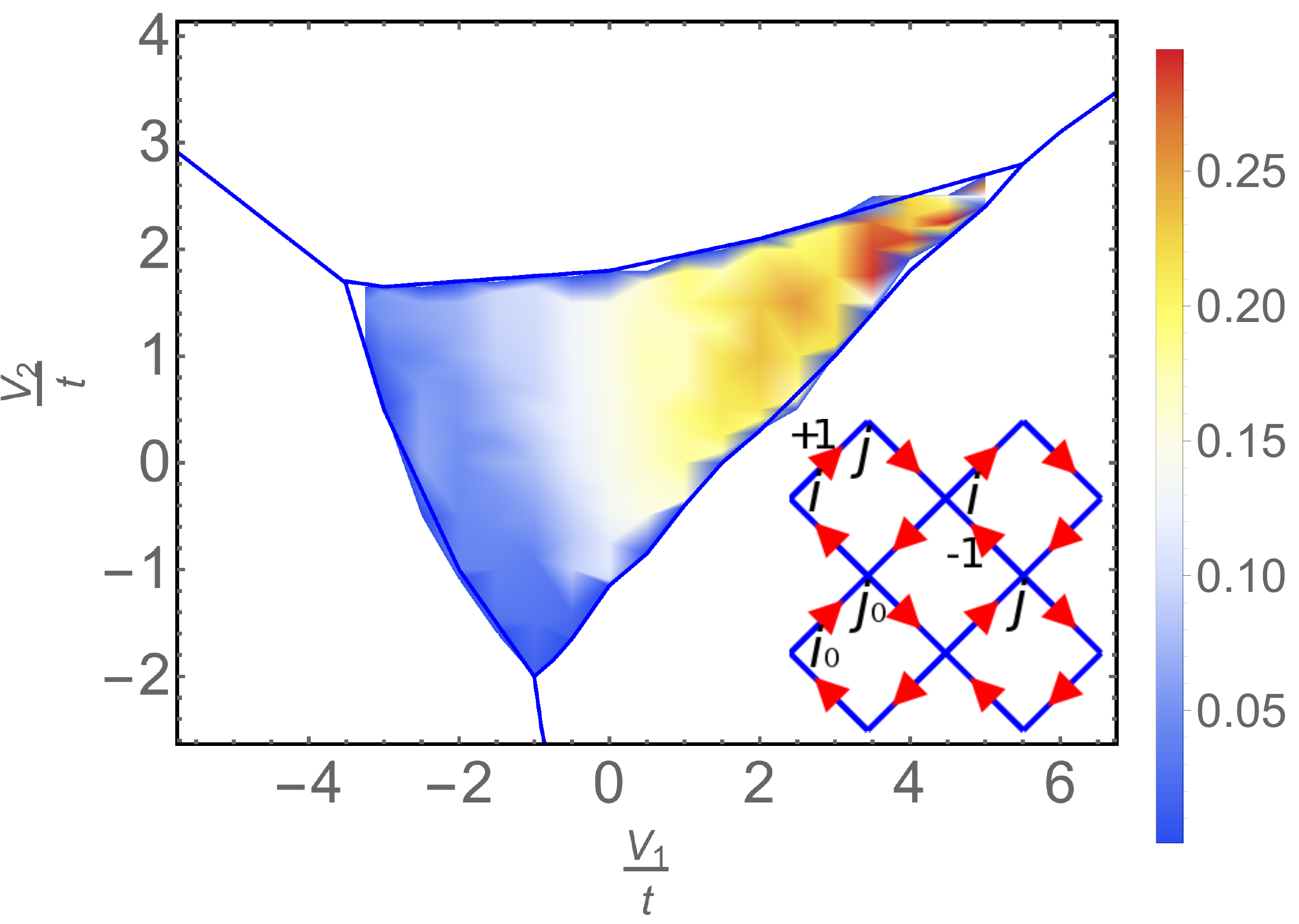}
\caption{Interaction dependence of the QAH structure factor.
The structure factor $\mathcal J_{\rm QAH}$ is calculated by the summation of the QAH current 
correlations $\langle \Delta_{ij} \Delta_{i_0 j_0} \rangle$ in the bulk of 
the cylinder as defined in Eq.~\eqref{eq:stru}. The data are obtained from 
the middle $4 \times 4$ unit cells on the $L_y = 4$ cylinder. 
The inset shows the sign convention for $\epsilon_{ij}$ in Eq.~\eqref{eq:stru}. For 
the reference bond $(i_0, j_0)$ with $i_0 \rightarrow j_0$ following the arrow 
direction, the bonds $(i, j)$ with the direction $i \rightarrow j$ following 
the arrow direction have $\epsilon_{ij} = 1$; otherwise if $i \rightarrow j$ 
follows the opposite direction, $\epsilon_{ij} = -1$.
}
\label{fig:structure}
\end{figure}

On the checkerboard lattice, we define the QAH order parameter as $\Delta_{ij} \equiv 4i \langle \Psi | c^{\dagger}_i c_j - c^{\dagger}_j c_i | \Psi \rangle$, where $| \Psi \rangle$ is the ground-state wavefunction and $i,j$ denote the sites connected by the nearest-neighbor bond. 
A nonzero $\Delta_{ij}$ implies  a spontaneously broken  time-reversal symmetry.
To obtain a global picture of the interaction dependence of the QAH order, we first calculate the QAH structure factor $\mathcal J_{\rm QAH}$ which is defined as a staggered sum of the current correlations $\langle \Delta_{ij} \Delta_{i_0 j_0} \rangle$,
\begin{equation}\label{eq:stru}
\mathcal J_{\rm QAH} = \frac{1}{N_s}\sum_{\langle ij \rangle} 
\epsilon_{ij} \langle \Delta_{ij} \Delta_{i_0 j_0} \rangle,
\end{equation}
where the sum runs over the nearest-neighbor bonds in the bulk of the cylinder (here we choose the middle $L_y \times L_y$ unit cells).
$N_s$ is the total number of the summed bonds, and $\epsilon_{ij} = \pm 1$ corresponds to the expected QAH current orientation of bond $(i, j)$ with respect to the reference bond, $(i_0, j_0)$. 
We show the current orientation of the QAH state in the inset of Fig.~\ref{fig:structure}, where $\Delta_{ij}$ is positive along the direction of the arrows.
In order for $\mathcal J_{\rm QAH}$ to be non-trivial, we take a reference bond $(i_0, j_0)$ in the bulk of the cylinder with $i_0 \rightarrow j_0$ following the arrow direction. 
Then if $i \rightarrow j$ follows the arrow direction we set $\epsilon_{ij} = + 1$;  otherwise  $\epsilon_{ij} = - 1$.
We show the structure factor on the $L_y = 4$ cylinder in Fig.~\ref{fig:structure}. 
In the region near $V_1 \sim V_2^2$, $\mathcal J_{\rm QAH}$  grows rapidly, which suggests a time-reversal symmetry breaking.

\begin{figure}[t]
\centering
\begin{subfigure}[b]{0.99\columnwidth}
\includegraphics[width=0.7\linewidth]{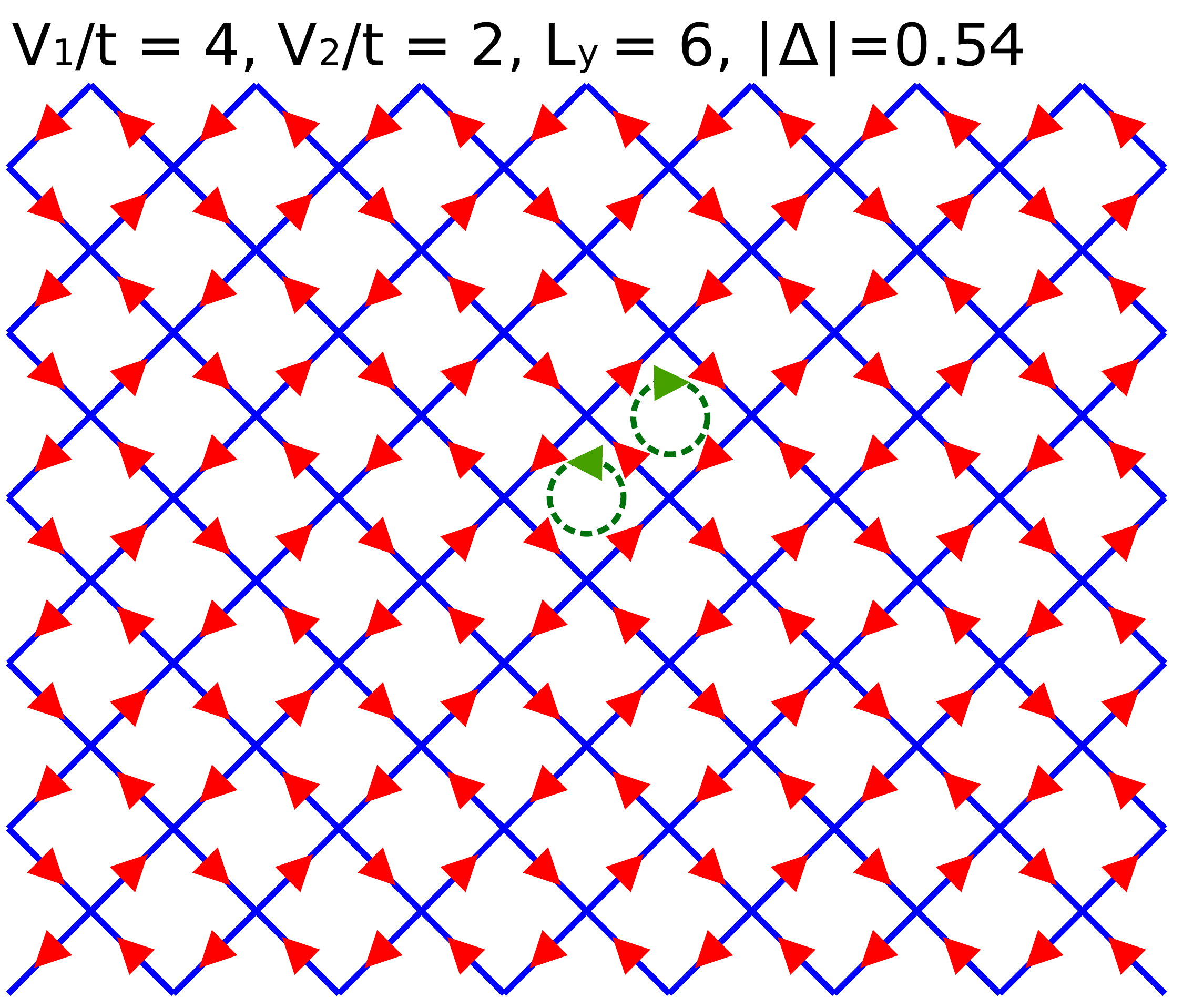}
\caption{}
\end{subfigure}
\begin{subfigure}[b]{0.99\columnwidth}
\includegraphics[width=0.9\linewidth]{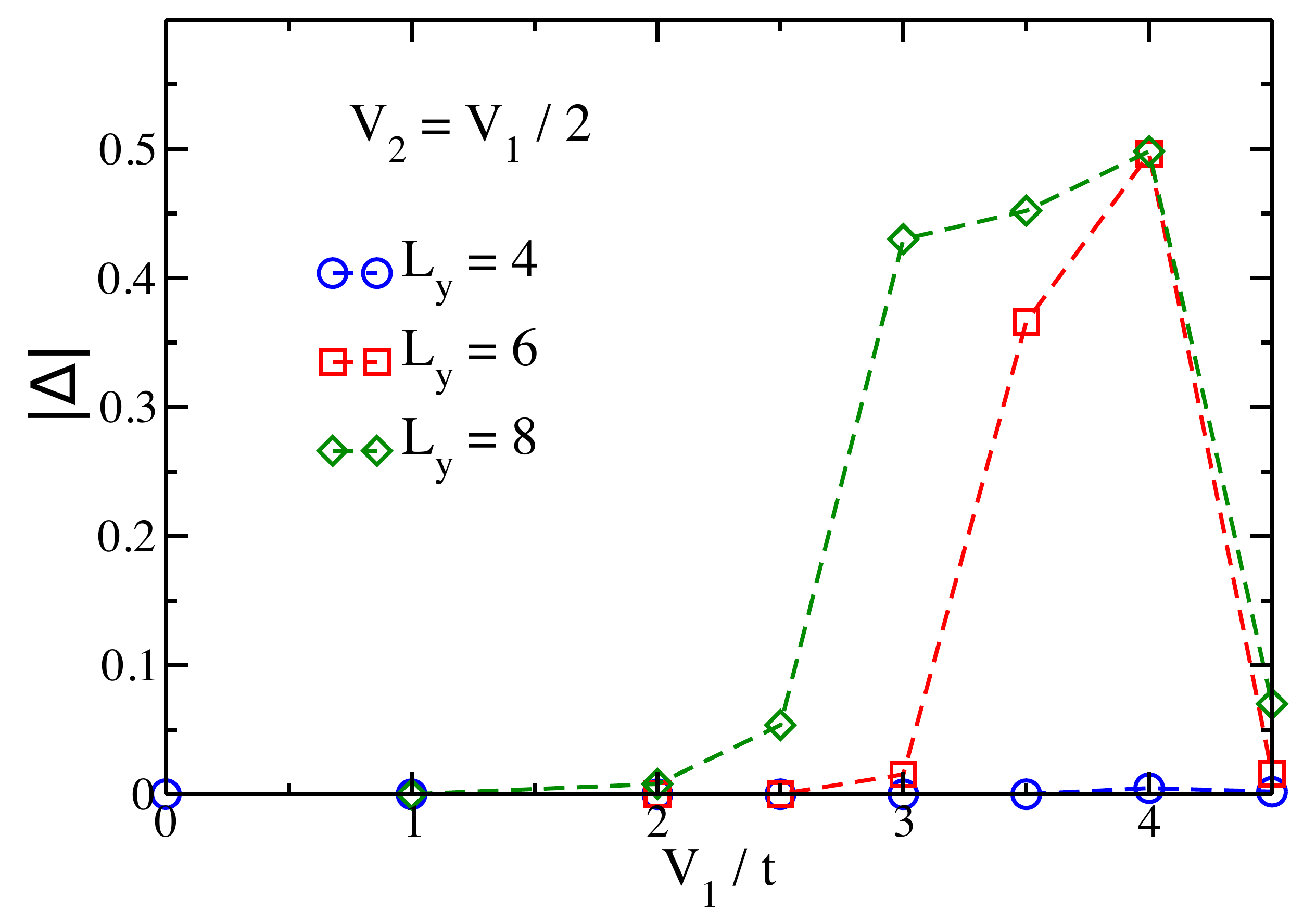}
\caption{}
\end{subfigure}
\caption{Spontaneous time-reversal symmetry breaking in the QAH state.
(a) Nonzero QAH order parameter $\Delta_{ij}$ for $V_1/t = 4, V_2/t = 2$ on the 
$L_y = 6$ cylinder. The arrow indicates that if the sites $i,j$ follow 
the arrow direction  $\Delta_{ij}$ is positive.
In the bulk of the cylinder, $\Delta_{ij}$ has a uniform value, $0.54$.
The green circle denotes the clockwise and counterclockwise directions in which  the 
loop-current circulates in each plaquette. 
The circulating loops have the opposite directions for the neighbor plaquettes, resulting in zero total flux.
(b) $V_1$ dependence of the QAH order parameter $\Delta$ for $V_1 = 2V_2$ on 
the $L_y = 4,6,8$ cylinders. For $V_1/t \gtrsim 2$, DMRG calculation with 
$L_y = 8$ finds a nonzero QAH order.
}
\label{fig:current}
\end{figure}

Next, we directly calculate the QAH order parameter $\Delta_{ij}$.
We use complex number wavefunction in DMRG simulation, which allows for a spontaneous time-reversal symmetry breaking leading to a nonzero $\Delta_{ij}$. 
This method has been widely used to identify time-reversal symmetry broken states such as QAH state~\cite{zhu2016} and chiral spin liquid~\cite{gong2014} in DMRG simulation.
In Fig.~\ref{fig:current}(a), we show the obtained $\Delta_{ij}$ for $V_1/t = 4, V_2/t = 2$ on the $L_y = 6$ cylinder.
We find a finite $\Delta_{ij}$ with a uniform magnitude in the bulk of the cylinder, which implies a spontaneously broken  time-reversal symmetry.
The local ordering pattern results in a loop current circulating in each plaquette.
The neighboring plaquettes  have an opposite circulation  direction, which leads to a  vanishing net flux and, thus, precisely agrees with the expectation of the QAH effect~\cite{haldane1988}.
By using the complex number wavefunction, we find one of the two degenerate time-reversal symmetry breaking ground states with either ``left-hand'' or ``right-hand'' chirality is spontaneously chosen.
The two states have the same energy but opposite QAH order.

To find the  region in the phase diagram where the ground state explicitly breaks  time-reversal symmetry, we measure the QAH order in the central triangular region in \fig{fig:phase}.
As the magnitude of the QAH order $|\Delta_{ij}|$ is uniform in the bulk of the cylinder, we simply denote it as $\Delta$. 
Here, we show the results along the line with $V_1/t = 2 V_2/t$ in Fig.~\ref{fig:current}(b) as a demonstrative example. 
On the $L_y = 4$ cylinder, $\Delta$ is vanishingly small for weak $V_1$, but obtains a finite value in the neighborhood of $V_1/t = 4, V_2/t = 2$. 
For $L_y = 6$, $\Delta$ at  $V_1/t = 3.5 \sim 4$ enhances dramatically (considered as a function of $L_y$).
The trend continues for $L_y = 8$, and $\Delta$ around $V_1/t = 4$ strengthens with increasing $L_y$ which indicates the presence of a robust time-reversal symmetry breaking. 
The small $\Delta$ around $V_1/t = 3$ on the $L_y = 6$ cylinder increases rapidly, showing that the a larger system-size overcomes the finite-size effects. 
Based on the results on the $L_y = 8$ cylinder, we find nonzero QAH order in the shaded region shown in Fig.~\ref{fig:phase}.

\subsection{Quantized Hall conductance} \label{sec:quantized-hall}

\begin{figure}[!t]
\centering
\begin{subfigure}[b]{0.99\columnwidth}
\includegraphics[width=0.99\columnwidth]{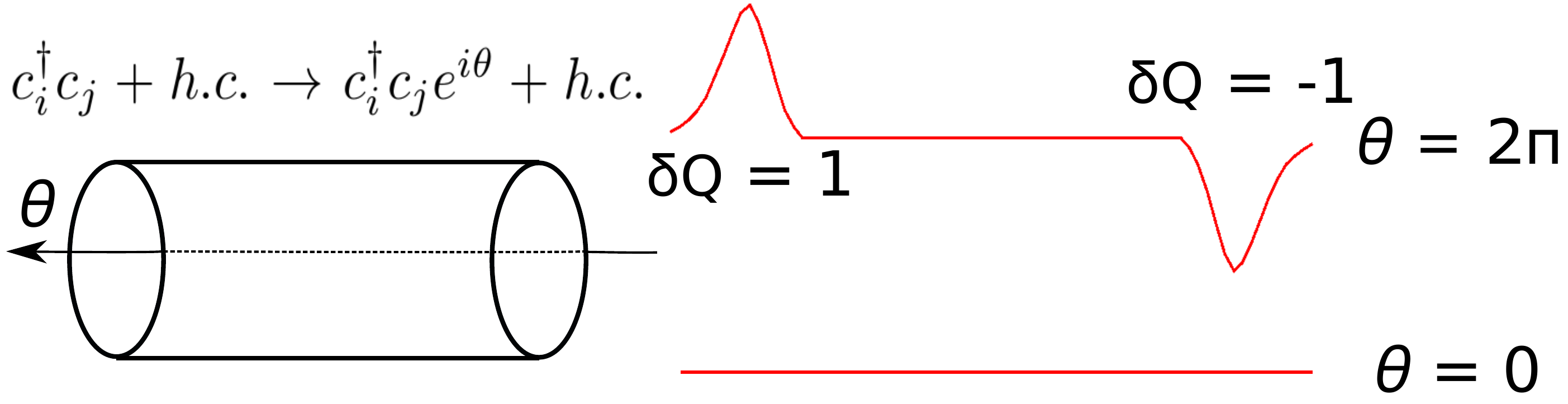}
\caption{}
\end{subfigure}
\begin{subfigure}[b]{0.99\columnwidth}
\includegraphics[width=0.9\columnwidth]{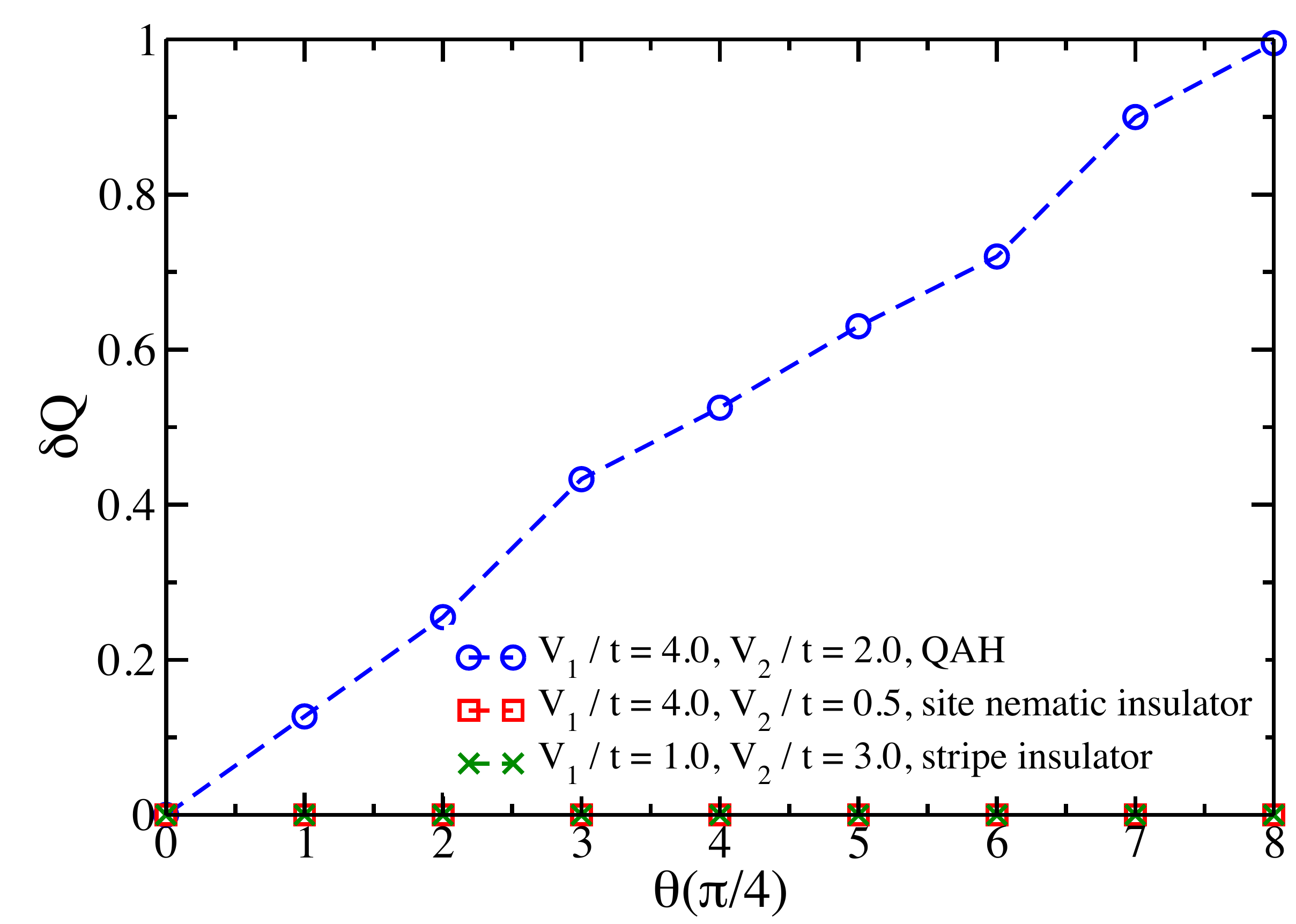}
\caption{}
\end{subfigure}
\caption{Charge pumping with inserting $U(1)$ flux in DMRG simulation.
(a) Schematic figure of the flux insertion simulation in DMRG. The 
charge flux $\theta$ is introduced in the cylinder by using   twisted 
boundary conditions for the hopping terms along the closed $y$ direction, 
i.e. for the hopping terms crossing the boundary line in the $y$ direction
we set $c^{\dagger}_i c_j e^{i\theta} + h.c.$. For an integer quantum 
Hall state, the charge will be pumped from one open edge of cylinder to 
the other edge by adiabatically increasing flux $\theta$. In a period of 
flux $\theta = 0 \rightarrow 2\pi$, a quantized charge $\delta Q$ will be transferred.
(b) Flux ($\theta$) dependence of the pumped charge number $\delta Q$ in the DMRG
simulation. We set the flux increase in units of  $\pi/4$. In the CDW phases, the charge
density $\langle n_i \rangle$ has no response to flux. In the QAH phase, 
the charge is pumped by  inserting flux. Over a period of $\theta$ 
a quantized net charge $\delta Q = 1$ is transfered, which characterizes the 
QAH phase as a Chern number $C = 1$ integer quantum Hall state.
}
\label{fig:flux}
\end{figure}

In order to reveal the topological nature of the QAH phase, we simulate the flux response in a  cylindrical system to measure the Hall conductance $\sigma_H$~\cite{gong2014kagome,zaletel2014}. 
Following the thought-experiment proposed by Laughlin for the integer quantum Hall  state~\cite{laughlin1981,sheng2003}, an integer quantized charge is expected to be pumped from one edge of the cylinder to the other  by inserting a period of $U(1)$ charge flux $\theta$ along the axis direction of the cylinder as shown in Fig.~\ref{fig:flux}(a). 
Over a period of flux $\theta = 0$ increases to $\theta= 2\pi$, the Hall conductance can be calculated from the pumped charge number $\delta Q$ with the help of  $\sigma_H = \frac{e^2}{h}\delta Q$~\cite{gong2014kagome,zaletel2014}.
In DMRG simulation, we introduce the charge flux by using the twisted boundary condition in the $y$ direction,  $c^{\dagger}_i c_j + h.c. \rightarrow c^{\dagger}_i c_j e^{i\theta} + h.c.$, for all the hopping terms that cross the $y$ boundary. 
With growing flux $\theta$, we use the adiabatic DMRG simulation by taking the converged ground state with a  given flux $\theta^{\prime}$ as the initial ground state for the next-step sweeping with the increased flux $\theta^{\prime} + \delta \theta$~\cite{gong2014kagome}.

By adiabatically inserting flux $\theta$ in DMRG simulation, we calculate the distribution of the charge density, $\langle n_i \rangle$, on the cylinder. 
In the charge ordered phases, the charge density has no response to flux as shown in Fig.~\ref{fig:flux}(b). 
In the parameter region with spontaneous time-reversal symmetry breaking, we find that the charge is pumped from one edge of the cylinder to the other without accumulation or depletion of the net charge in the bulk of the cylinder, i.e. the charge density of the sites in the bulk of the cylinder is always $1/2$ during the whole pumping process.
In a period of flux $\theta = 0 \rightarrow 2\pi$, the pumped net charge $\delta Q = 1.0$, which characterizes the quantized Hall conductance and identifies the QAH phase as a Chern number $C = 1$ integer quantum Hall phase.

\subsection{Decay length of the QAH order parameter}

\begin{figure}[t]
\centering
\begin{subfigure}[b]{0.99\columnwidth}
\includegraphics[width=0.6\linewidth]{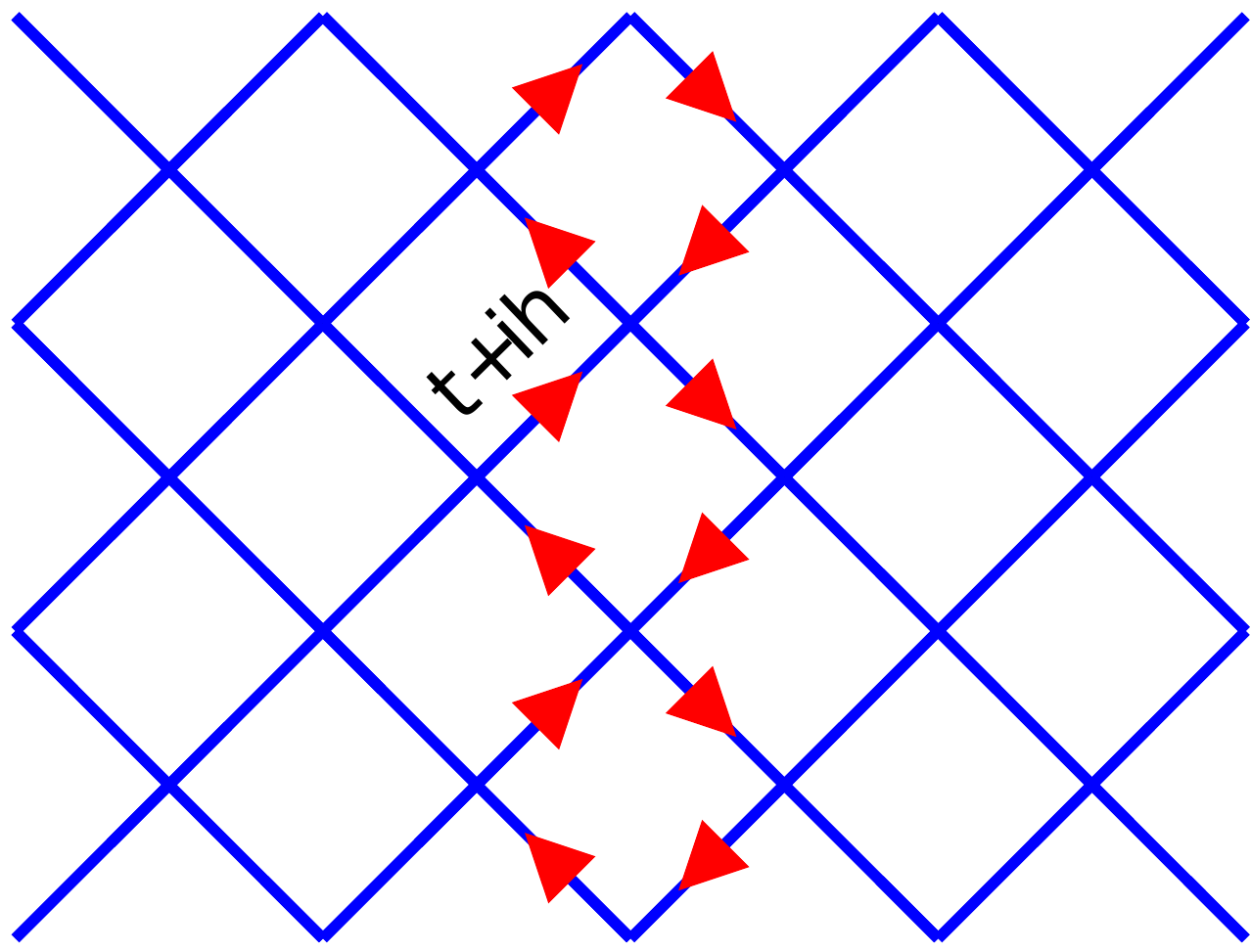}
\caption{}
\end{subfigure}
\begin{subfigure}[b]{0.99\columnwidth}
\includegraphics[width=1.0\linewidth]{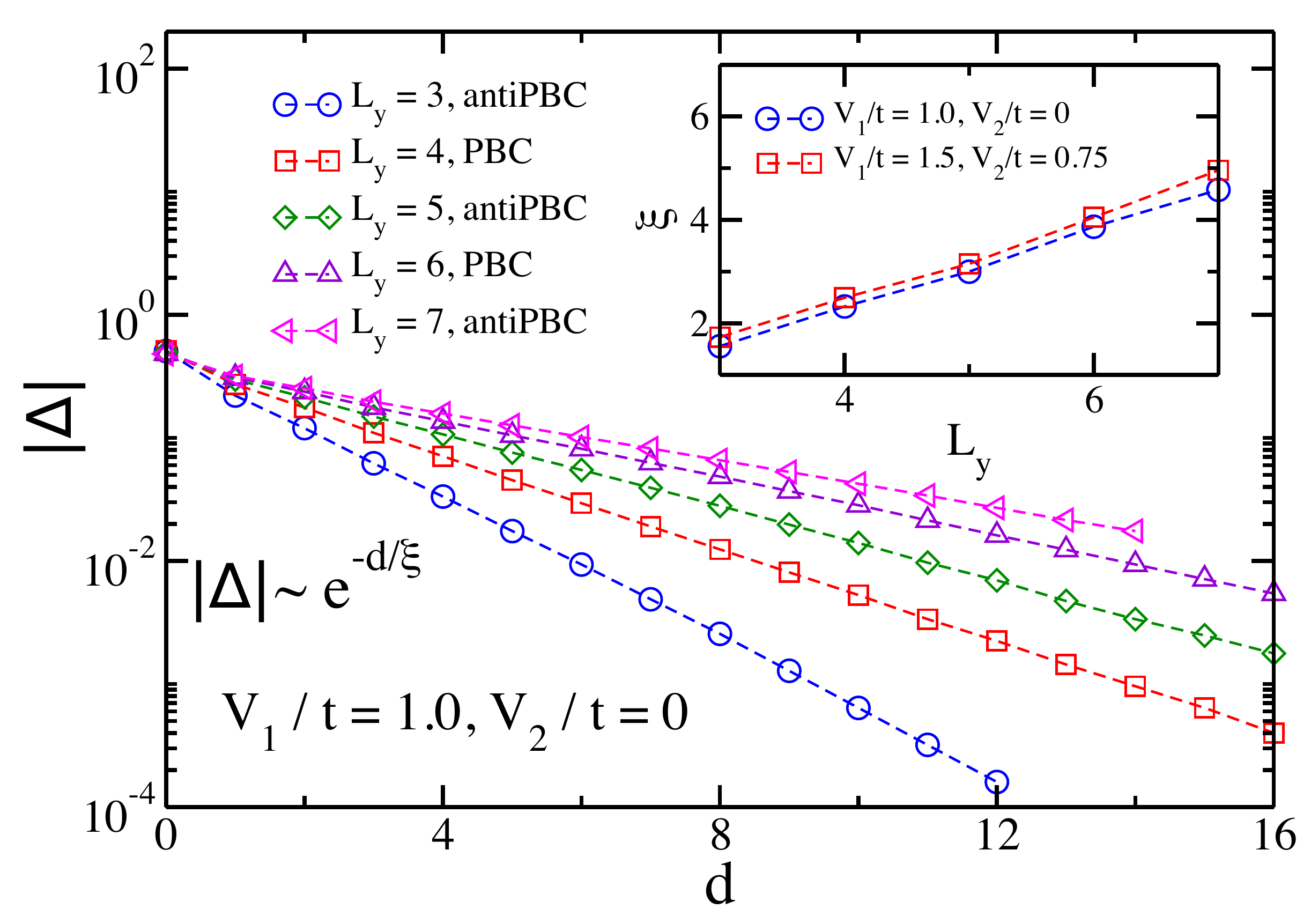}
\caption{}
\end{subfigure}
\caption{Decay length of the QAH order parameter which is driven by a pinning field.
(a) Schematic figure of the QAH pinning field. In the middle column of the long cylinder, 
the hoppings of the nearest-neighbor bonds with the red arrows are changed from 
$t c^{\dagger}_i c_j + h.c.$ to $(t+ih) c^{\dagger}_i c_j + h.c.$, where $i \rightarrow j$
follows the direction of the arrow. $h$ is the pinning field, which breaks time-reversal
symmetry and leads to a nonzero QAH order, $\Delta$, that decays from the pinning column 
to the edge.
(b) Log-linear plot of the QAH order driven by the pinning field versus the distance of the
measured bond to the pinning column. The system has $V_1 / t = 1.0, V_2 / t  = 0.0$. 
The even $L_y$ cylinder with  periodic boundary condition  and the odd $L_y$ cylinder with the anti-periodic boundary condition  are studied. 
The QAH order $\Delta$ decays exponentially from the middle column to the edge, giving a 
decay length $\xi$ from $\Delta \sim e^{-d/\xi}$. The inset shows the $L_y$ dependence of $\xi$,
where $\xi$ grows almost linearly with increasing $L_y$.}
\label{fig:chiraldecay}
\end{figure}

In the parameter regime where the interaction is repulsive and our DMRG simulation does not find an  unambiguous evidence for a QAH phase, we measure the decay length of the QAH order parameter by adding a pinning field in the bulk of the cylinder.
We introduce a pinning field in a single column of bonds in the middle of the cylinder by modifying the nearest-neighbor hopping from $t c^{\dagger}_i c_j + h.c.$ to $(t+ih) c^{\dagger}_i c_j + h.c.$, where $h$ is the pinning field which follows the direction shown in Fig.~\ref{fig:chiraldecay}(a).
Since a finite $h$ breaks time-reversal symmetry, $\Delta$ obtains a finite value on the pinning bonds.
The nonzero QAH order exponentially decays along the $x$ direction as $\Delta \sim e^{-d/\xi}$, where $d$ is the distance of the measured bond from the pinning column, and  $\xi$ is the decay length. 
For a system that is too small for an unambiguous  detection of the QAH order with the methods discussed in Sections \ref{sec:TRS} and \ref{sec:quantized-hall}, we may still identify the QAH order by measuring how the decay length $\xi$ scales with increasing $L_y$.
If $\xi$ diverges with  $L_y$ then QAH is realized in a sufficiently large system.
In contrast, if $\xi$ approaches a finite value in the large $L_y$ limit, then the QAH order is absent.
This method has been successfully used to detect the weak valence bond order in quantum spin systems~\cite{sandvik2012,zhu2013,gong2013,gong2014}. 
We first test the system with $V_1/t = 1.0, V_2/t = 0.0$ on  even $L_y$ cylinder with the periodic boundary condition, and odd $L_y$ cylinder with the anti-periodic boundary condition   
\footnote{For these boundary conditions the QBT is present in the non-interacting single-particle dispersion along the $\hat x$ direction.}.
In Fig.~\ref{fig:chiraldecay}(b), we show the log-linear plot of the QAH order $\Delta$ versus  $d$. 
As anticipated, $\Delta$ decays exponentially, and the decay length, $\xi$, is shown in the inset. 
In our simulation we find that although $\Delta$ depends on the pinning field strength, the decay length is almost independent of $h$, which has also been found in the dimer pinning~\cite{gong2014}. 
On the $V_1$ axis $\xi$ grows with $L_y$ and does not show any saturation.
A similar behavior is also found away from the $V_1$ axis in the presence of a repulsive $V_2$.
In the inset of Fig.~\ref{fig:chiraldecay}(b) we demonstrate this behavior at two sample points in the phase diagram.
The fast increase of decay length with $L_y$ is consistent with the presence of a QAH phase.
Therefore, our DMRG simulation fully establishes a QAH phase over a large region in the  $V_1 - V_2$ phase diagram.

\section{Power expanded Gibbs potential analysis} \label{sec:pegp}
\begin{figure}[!t]
\centering
\includegraphics[width=0.5\columnwidth]{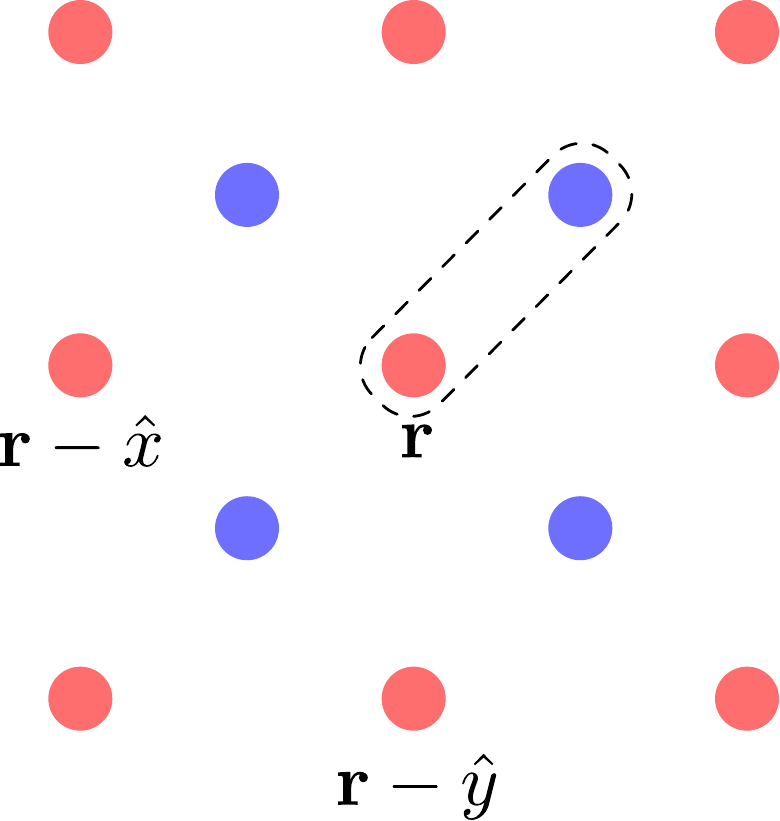}
\caption{The checkerboard lattice is considered as a decorated square lattice. The $\A$-sites (red) occupy the lattice points of the square lattice, while the $\B$-sites (blue) are displaced by $(\hat{x} + \hat{y})/2$ with respect to the $\A$-site of the same unit cell. The dashed square represents the unit cell at $\bs r$.}
\label{fig:lattice-scheme}
\end{figure}

In this section we utilize the power expanded Gibbs potential method (PEGP) for calculating the QAH order as a function of the couplings $V_1$ and $V_2$. 
The PEGP was introduced in the study of spin glass order in the infinite-ranged Ising model below the critical temperature~\cite{plefka1982}. 
Here we adopt this method for the analysis of the  zero-temperature phase diagram.
The main advantage of PEGP over conventional mean-field theory is its ability to track orderings that result entirely through quantum fluctuations,  including those that cannot be obtained by a mean-field decomposition of the terms in the classical theory.
In the present model, under coarse-graining the next-nearest-neighbor interaction generates an effective nearest-neighbor interaction which in turn drives the weak-coupling instability of the QBT semimetal.
The PEGP precisely captures this process, and yields the dependence of the QAH order on $V_1$ and $V_2$.
Although our analytical computation focuses on the  weak-coupling region, in principle, the method can be used to explore the phase diagram beyond strict weak coupling regime.
We consider the checkerboard lattice as a decorated square lattice with two sites, $\A$ and $\B$,  per unit cell as illustrated in \fig{fig:lattice-scheme}.
This leads to both inter-unit cell and intra-unit cell hoppings and repulsive interactions,
\begin{widetext}
\begin{align}
H &= \sum_{\mbf r} \lt[ \A_{\mbf r}^{\dag}  \B_{\mbf r} 
+  \A_{\mbf r}^{\dag}  \B_{\mbf r - \hat{x}} 
+ \A_{\mbf r}^{\dag}  \B_{\mbf r - \hat{y}} 
+ \A_{\mbf r}^{\dag}  \B_{\mbf r - \hat{x} - \hat{y}} + \mbox{h.c.}\rt] 
+ \half \sum_{\mbf r} \lt[ \A_{\mbf r}^{\dag}  \A_{\mbf r - \hat{y}} 
+  \B_{\mbf r - \hat{x}}^{\dag}  \B_{\mbf r} 
+ \mbox{h.c.} \rt] 
-\half \sum_{\mbf r} \lt[ \A_{\mbf r}^{\dag}  \A_{\mbf r - \hat{x}} 
+  \B_{\mbf r - \hat{y}}^{\dag}  \B_{\mbf r} 
+ \mbox{h.c.} \rt]
\nn \\
&  
+ V_1 \sum_{\mbf r} \A_{\mbf r}^{\dag} \A_{\mbf r} \lt( \B_{\mbf r}^{\dag} \B_{\mbf r} + \B_{\mbf r-\hat x}^{\dag} \B_{\mbf r - \hat x} 
+ \B_{\mbf r - \hat y}^{\dag} \B_{\mbf r - \hat y} 
+ \B_{\mbf r -\hat x - \hat y}^{\dag} \B_{\mbf r - \hat x - \hat y} \rt) \nn \\
&~ + V_2 \sum_{\mbf r} \lt[ \A_{\mbf r}^{\dag} \A_{\mbf r} \lt(\A_{\mbf r + \hat x}^{\dag} \A_{\mbf r + \hat x} 
+ \A_{\mbf r + \hat y}^{\dag} \A_{\mbf r + \hat y} 
+ \A_{\mbf r - \hat x}^{\dag} \A_{\mbf r - \hat x} 
+ \A_{\mbf r - \hat y}^{\dag} \A_{\mbf r - \hat y} \rt) + (\A \ltrtarw \B) \rt],
\label{eq:h}
\end{align}
\end{widetext}
where $\mbf r$ denotes the position of an unit cell.
We set the lattice spacing to unity and consider an infinite system to define the Fourier components,
\begin{align}
& \qty{\A_{\mbf r},\B_{\mbf r}} = \int \dd{\bs k}  e^{i \mbf r \cdot \bs k} ~ \qty{\A(\bs k), e^{(i/2) (\hat x + \hat y) \cdot \bs k} \B(\bs k)}
\label{eq:FT}
\end{align}
where $\bs k$ lies within the  first Brillouin zone, and $d\bs{k} \equiv \frac{dk_x dk_y}{(2\pi)^2}$.
Therefore, the action in momentum space representation takes the form,
\begin{widetext}
\begin{align}
S &= \int \dd{k} \psi^{\dag}(k) \lt[ -ik_0 \sig_0 + d_1(\bs k) \sig_1 + d_3(\bs k) \sig_3 \rt] \psi(k) \nn \\
& + \int \dd{k} \dd{k'} \dd{q} 
\Bigl[
4 V_1 \cos{\frac{q_x}{2}} \cos{\frac{q_y}{2}} ~ \psi^{\dag}(k + q) \frac{\sig_0 + \sig_3}{2} \psi(k) \psi^{\dag}(k') \frac{\sig_0 - \sig_3}{2} \psi(k' +  q)  \nn \\
& \quad + 2V_2 (\cos{q_x} + \cos{q_y} -2 ) \sum_{s=\pm }\psi^{\dag}(k + q) \frac{\sig_0 + s \sig_3}{2} \psi(k) \psi^{\dag}(k') \frac{\sig_0 + s \sig_3}{2} \psi(k' +  q)   \Bigr],
\label{eq:S} 
\end{align}
\end{widetext}
where $\dd{k} \equiv \int_{-\infty}^{\infty} \frac{dk_0}{2\pi} \int d\bs{k}$, $\psi(k) = \trans{(\mf{a}(k), \mf{b}(k))}$ is a two-component Grassman spinor, $d_1(\bs k) = 4 \cos{\frac{k_x}{2}} \cos{\frac{k_y}{2}}$, $d_2(\bs k) = 4 \sin{\frac{k_x}{2}} \sin{\frac{k_y}{2}}$, and $d_3(\bs k) =  \cos{k_x} -  \cos{k_y}$, $\sig_0$ is the $2\times 2$ identity matrix, and $\sig_i$ are the Pauli matrices.

We express the local QAH order parameter as (see \fig{fig:lattice-scheme}),
\begin{align} \label{eq:qah}
\Dl(\bs r) &\equiv \gam(\bs r; 0) - \gam(\bs r; \hat x) - \gam(\bs r; \hat y) + \gam(\bs r; \hat{x} + \hat{y}),
\end{align}
where  
\begin{align}
\gam(\bs r; \bs w) = i \lt[ a_{\bs r}^{\dag} b_{\bs r - \bs w} - b_{\bs r - \bs w}^{\dag} a_{\bs r}\rt].
\end{align}
In Ref.~\cite{sun2009} the authors have studied the model, Eq.~\eqref{eq:S},  
in the absence of the $V_2$ term, and established a mean-field phase diagram where the QAH order parameter, $\Dl(\bs r) \sim {\Lam_0}^2 e^{-1/V_1}$ with ${\Lam_0}$ being an effective momentum scale.
While the QAH phase is stabilized over a larger region of the phase diagram in the presence of the $V_2$ term as established by our DMRG calculation, it is not possible to show this within a conventional mean-field theoretic framework. 
The main obstruction results from $\Dl(\bs r)$ not being obtainable by a mean-field decomposition of the $V_2$ vertex \footnote{In Ref. \cite{raghu2008} the QAH order on the honeycomb lattice is defined on the second-neighbor bond, which allows for a conventional mean-field analysis.}.
Moreover, the $V_2$ term is irrelevant in RG sense because it scales as $|\bs q|^2$ close to the $\bs M = (\pi, \pi)$ point, and nominally cannot drive a phase transition at weak coupling.
It is, however, a dangerously irrelevant operator, since its quantum fluctuation generates the marginally relevant operator that destabilizes the QBT semimetal.
Thus, in order to study the QAH phase on the $V_1 - V_2$ plane we utilize the PEGP which does not rely on explicit mean-field decoupling of the interaction vertices.
In the following subsections we outline the general principles of PEGP, and then use it to deduce the phase diagram.

\subsection{General formalism}
Here we briefly review the PEGP formalism for a system of finite size and at finite temperature~\cite{plefka1982}.
We extend the Hamiltonian in \eq{eq:h} by introducing an artificial parameter, $\pha$, and a source, $J$, for the order parameter of interest,  $\mathcal{O}$, and schematically express it as,
\begin{equation}\label{eq:newh}
H[\pha, J] = H_0 + \alpha H_{\rm int} + J\mathcal{O},
\end{equation}
where $H_0$ is the non-interacting Hamiltonian, and $H_{\rm int}$ is the interaction term.
We note that $\alpha = 1$ corresponds to \eq{eq:h} in the presence of the source term. 
The Gibbs potential is given by
\begin{equation}\label{eq:freeenergy}
\mathcal{G}(\alpha,\beta,\Delta)=-\frac{1}{\beta}\ln[{\rm Tr}(e^{-\beta (H_0+\alpha H_{\rm int}+J\mathcal{O})})]
-L^2 J\Delta,
\end{equation}
where $L$ is the system size and $\Dl = \langle \mathcal{O}\rangle_{\alpha}/ L^2$. 
For the QAH order, $\Delta$ is given by Eq.~\eqref{eq:qah}. 
Here $\langle \cdots\rangle_{\alpha}$ denotes the expectation value with respect to  $H[\pha, J]$.
We note that in the Gibbs potential the order parameter $\Delta$ is an independent variable, and  $J$ is a function of $\alpha, \beta$ 
and $\Delta$ which, in principal, can be obtained by inverting the relation 
$L^2 \Delta = \langle \mathcal{O}\rangle_{\alpha}$.

The Gibbs potential is computed perturbatively by expanding it  in powers of $\alpha$ around $\pha = 0$,
\begin{align}
 \mathcal{G}(\alpha,\beta,\Delta) &= \mathcal{G}(0,\beta,\Delta) 
+ \lt\{\lt. \frac{\partial \mathcal{G}(\alpha,\beta,\Delta)}{\partial \alpha}\rt|_{\alpha = 0} \rt\} \alpha \nonumber \\
& + \frac{1}{2} \lt\{\lt. \frac{\partial^2 \mathcal{G}(\alpha,\beta,\Delta)}{\partial^2 \alpha}\rt|_{\alpha = 0} \rt\} \alpha^2 
+ \ordr{\alpha^3}.
\label{eq:expansion}
\end{align}
In the weak coupling limit we can truncate the expansion at quadratic order, and take  $\pha \rtarw 1$ to obtain, 
\begin{widetext}
\begin{align}
\mathcal{G}(1,\beta,\Delta) &\simeq \mathcal{G}(0,\beta,\Delta) + \langle H_{\rm int}\rangle_0
+ \frac{\beta}{2} 
\lt(
\langle H_{\rm int} \rangle^2_0 - \langle H^2_{\rm int} \rangle_0 
+ \frac{\partial J}{\partial \alpha} \langle \mathcal{O} \rangle_0 \langle H_{\rm int} \rangle_0
- \frac{\partial J}{\partial \alpha} \langle H_{\rm int} \mathcal{O} \rangle_0
\rt),
\label{eq:freeenergy2}
\end{align}
\end{widetext}
where we have used the relations, 
\begin{align}
& \frac{\partial \mathcal{G}(\alpha,\beta,\Delta)}{\partial \alpha} = \langle H_{\rm int}\rangle_{\alpha}, \\
& \frac{\partial^2 \mathcal{G}(\alpha,\beta,\Delta)}{\partial^2 \alpha} = 
\beta \langle H_{\rm int} \rangle^2_{\alpha} 
+ \beta \frac{\partial J}{\partial \alpha} \langle \mathcal{O} \rangle_{\alpha} 
\langle H_{\rm int} \rangle_{\alpha} \nonumber \\
&\hspace{0.3\columnwidth} - \beta \langle H^2_{\rm int} \rangle_{\alpha} 
- \beta \frac{\partial J}{\partial \alpha} \langle H_{\rm int} \mathcal{O} \rangle_{\alpha},
\end{align}
and the thermodynamic relation 
${\partial \mathcal{G}}/{\partial \Delta} = -L^2 J$.
From the roots of the equation  $\partial \mathcal{G} / \partial \Delta = 0$ we determine the dependence of $J$ on the couplings with the help of the chain rule,  
$\partial \mathcal{G} / \partial \Dl = (\partial \mathcal{G} / \partial J) (\partial J / \partial \Delta)$.
Here $\partial \mathcal{G} / \partial J$ is obtained from Eq.~\eqref{eq:freeenergy2}, while  $\partial J / \partial \Delta$ is calculated by inverting the relation
$L^2 \Delta = \langle \mathcal{O}\rangle$. 
The expression of the order parameter, $\Delta$, that minimizes $\mathcal{G}$ is in turn obtained by using the relationship between $J$ and $\Dl$.
In the following subsection we demonstrate the method for an effective continuum model that follows from \eq{eq:S}.


\subsection{PEGP analysis of the effective low energy theory} \label{sec:eff-S}
\begin{figure}[!t]
\centering
\includegraphics[width=0.7\columnwidth]{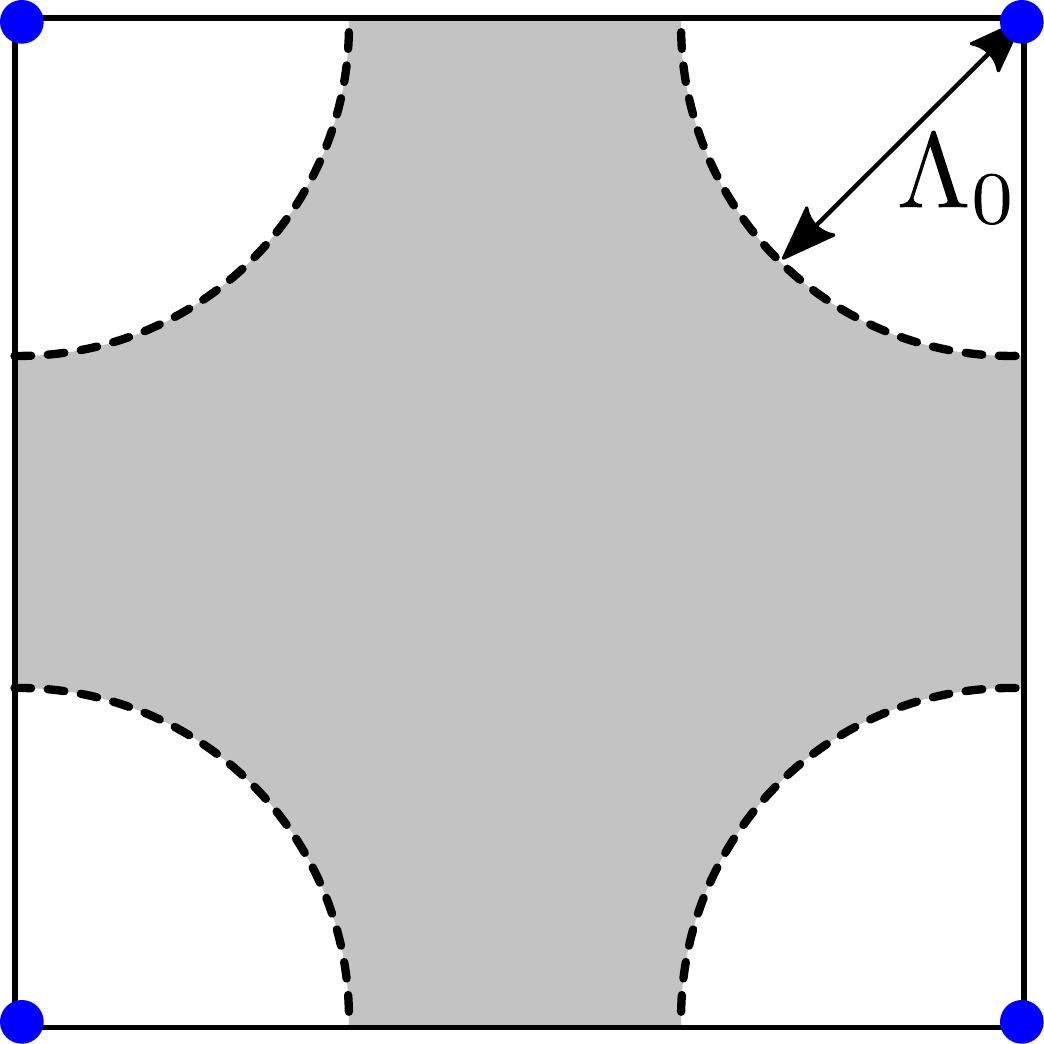}
\caption{Construction of the  effective theory. The square represents the 1st Brillouin zone. The filled (blue) circles are equivalent due to Brillouin zone periodicity and host the QBT. The modes in the shaded region are integrated out to obtain the low energy effective action in \eq{eq:eff-S} defined with the UV cutoff $\Lam_0$.}
\label{fig:coarse-grain}
\end{figure}
\begin{figure}[!t]
\centering
\includegraphics[width=0.5\columnwidth]{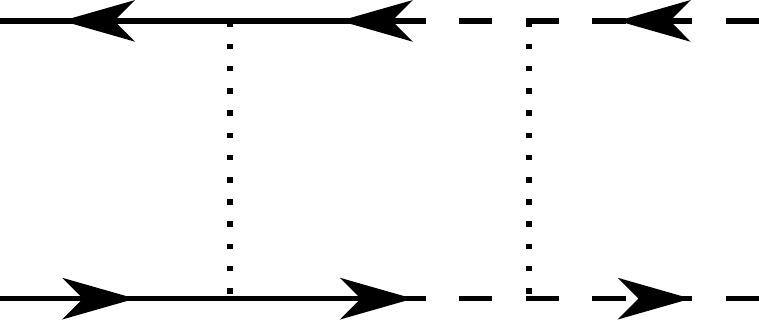}
\caption{The quantum fluctuation at order $V_2^2$ that generates the marginal interaction vertex in the low energy effective theory. The solid (dashed) lines represent  $\mf a$ ($\mf b$) type fermion, and the dotted line is the momentum dependent coupling function in \eq{eq:S}}
\label{fig:ph}
\end{figure}

In this section we use the PEGP formalism to obtain an expression of the QAH order from a low energy effective theory in the thermodynamic limit with $T=0$.
Since the logarithms that lead to QAH instabilities 
result from the infrared (IR) sector, the low energy effective theory is expected to be sufficient for obtaining 
qualitatively correct results.

We focus on a small neighborhood of radius ${\Lam_0}$ centered at the QBT at $\bs{M} = (\pi, \pi)$, with ${\Lam_0} \ll 1$ in units of inverse lattice spacing.
In order to obtain the effective action we expand the dispersion and the coupling functions around $\bs M$.
Although the $V_2$ vertex is suppressed by a factor of $|\bs q|^2$ in the low energy limit, it renormalizes the $V_1$ vertex through quantum fluctuations.
Therefore, the bare strength of the marginal interaction in the effective theory is controlled by both $V_1$ and $V_2$.
In order to obtain an expression for the bare value of this effective coupling, we integrate out modes that lie in the shaded region in \fig{fig:coarse-grain}. 
We assume $V_1$ to be sufficiently weak such that modes above ${\Lam_0}$ coupled through $V_1$ do not lead to significant renormalizations.
The effective action takes the form,
\begin{align}
& S = \int_{{\Lam_0}} dk ~ \Psi^\dag(k) \lt[ik_0 \sigma_0 + \mc{E}_0(\bs{k}) \rt] \Psi(k) \nn \\
& + g(\Lam_0) \int_{\Lam_0} \lt(\prod_{n=1}^4 dk_n \rt) \psi_a^\dag(k_1) \psi_a(k_2) \psi_b^\dag(k_3)  \psi_b(k_4),
\label{eq:eff-S}
\end{align}
where $\int_{\Lam_0}$ implies  $|\bs{k}|<{\Lam_0}$, 
$\Psi = (\psi_a, \psi_b)^{\intercal}$ with $\{\psi_a, \psi_b\}$ being the coarse-grained modes carrying momenta around $\bs{M}$,  $\mc{E}_0(\bs k)$ is the 
dispersion in the neighborhood of $\bs M$,
\begin{align}
\mc{E}_0(\bs{k}) = \frac{1}{2} |\bs{k}|^2 \sin{2\theta_k} ~\sigma_1 + \frac{1}{2} |\bs{k}|^2 \cos{2\theta_k} ~\sigma_3 
+ \ordr{|\bs k|^4},
\label{eq:disp}
\end{align}
with $\theta_k$ being the angular position of $\bs k$, with respect to $\bs M$, and 
\begin{align}
g({\Lam_0}) = 4 V_1 + \alpha(\Lam_0) V_2^2 
\label{eq:eff-g}
\end{align} 
is the effective coupling at the UV scale, ${\Lam_0}$.
The $V_2^2$ term is generated by the quantum fluctuation in \fig{fig:ph} \footnote{The other two one-loop diagrams in the particle-hole channel generate irrelevant effective vertices which we ignore.}.
In order to simplify the analysis, henceforth we replace $\alpha(\Lam_0)$ by the limiting value, $\alpha(\Lam_0 \rtarw 0) = 1.45$, such that $g(\Lam_0) \rtarw g_0 = 4 V_1 + 1.45 V_2^2$.
We note that we have ignored renormalizations to the quadratic part of the action.
The asymptotic behavior of \eq{eq:eff-S} was studied in Ref.~\cite{sun2009}.
In particular, $g$ was shown to be marginally relevant, and within a mean-field analysis it was shown to drive the system into a QAH state.
We note that the microscopic model that led to the effective action in Ref.~\cite{sun2009} corresponds to the $V_2 = 0$ limit of our model.

\begin{figure}[!t]
\centering
\begin{subfigure}[b]{0.48\columnwidth}
\includegraphics[width=0.9\columnwidth]{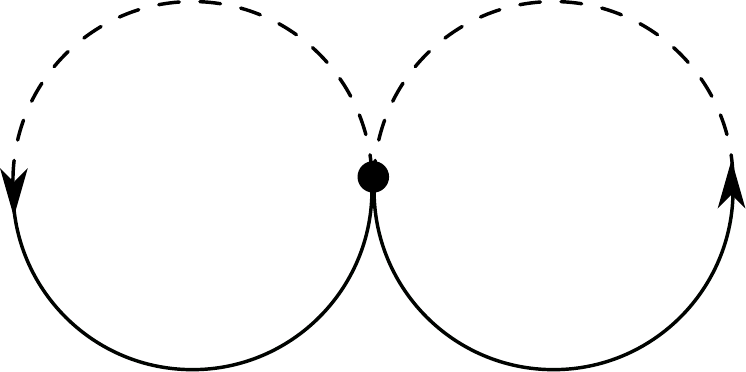}
\caption{}
\label{fig:order-1-1}
\end{subfigure}
\hfill
\begin{subfigure}[b]{0.48\columnwidth}
\includegraphics[width=0.9\columnwidth]{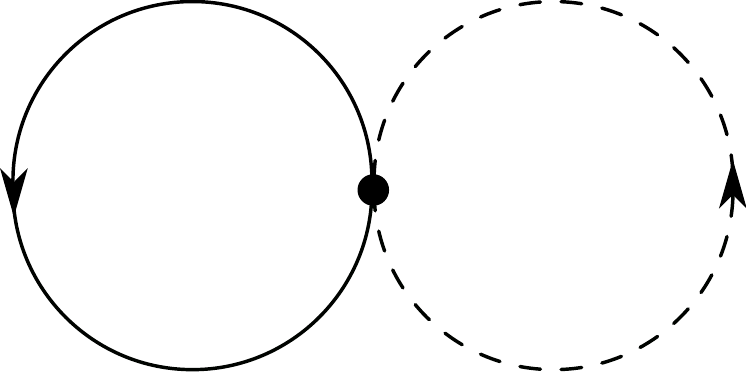}
\caption{}
\label{fig:order-1-2}
\end{subfigure}
\caption{Vacuum diagrams at the linear order in $g_0$. The solid (dashed) lines represent  $\mf a$ ($\mf b$) type fermion. The mixed lines represent the off-diagonal terms in the matrix propagator. Here the coupling function is momentum independent (i.e. a constant, $g_0$) and represented by the filled circle. }
\label{fig:order-1}
\end{figure}

We demonstrate the PEGP method with the help of \eq{eq:eff-S}, and derive an expression for the QAH order parameter which \emph{implicitly} depends on $V_1$ and $V_2$ through the bare effective coupling, $g_0$.
We introduce a source, $J$, for the QAH state which amounts to addition of the term, $\half J \int dk \Psi^{\dag}(k) (8 - |\bs{k}|^2 ) ~ \sigma_2 \Psi(k)$ to the effective action.
Thus the propagator in the presence of the source is 
\begin{align}
G(k; J)^{-1} = i k_0 \sig_0 + \mc{E}_0(\bs k) + \half J (8 - |\bs{k}|^2 ) ~ \sigma_2.
\end{align}
The Gibbs potential up to linear order in $g_0$ is
\begin{align}
\mathcal G(g_0, \Delta) = \mathcal G(0, \Delta) + \langle S_{\rm int}\rangle_{J}.
\end{align}
Two different processes contribute to $\langle S_{\rm int}\rangle_{J}$, as shown in Fig.~\ref{fig:order-1}.
While the process in \fig{fig:order-1-2} averages to 0, \fig{fig:order-1-1} leads to a nonzero contribution,
\begin{align}
& \langle S_{\rm int}\rangle_{J} =  - \frac{g_0 J^2}{4} 
\left( \int_{\Lam_0} dk  \frac{(8-|\bs{k}|^2)}
{k_0^2 + \frac{1}{4} |\bs{k}|^4 + \frac{J^2}{4} (8 - |\bs{k}|^2 )^2 } \right)^2.
\end{align}
Therefore, retaining terms that do not vanish in the $\sqrt{J}/{\Lam_0} \rtarw 0$ limit, we obtain 
\begin{align}
\partial_{\Delta} \mathcal G(g_0, \Delta) = - J - A_0 ~g_0 ~J \ln{\frac{J}{{\Lam_0}^2}},
\label{eq:soln-eff-S}
\end{align}
where $A_0 > 0$ is a numerical factor, and we have used the relationships,  $\partial_{\Delta} \mathcal G(0, \Delta) = -J$ and $\Delta(J) = - \half \int_{\Lam_0} dk (8 - |\bs{k}|^2 )  \tr{\sigma_2 G(k; J)}$.
It is straightforward to deduce that $\partial_{\Delta} \mathcal G(g_0, \Delta)$ 
vanishes at
\begin{align}
J = J^* \equiv \Lam_0^2 \exp\left\{-\frac{1}{A_0 g_0}\right\},
\end{align}
and  $\partial_{\Delta}^2 \mathcal G(g_0, \Delta(J^*)) > 0$.
Since $\Delta \sim J \ln{J}$, we obtain 
\begin{align}\label{eq:order}
\Delta \sim - \frac{\Lam_0^2}{A_0 g_0} \exp\left\{-\frac{1}{A_0 g_0}\right\}.
\end{align}
Therefore, on approaching the semimetallic phase from the ordered side $\Delta$ vanishes on the line, $4 V_1 + 1.45 V_2^2 = 0$, which identifies the phase boundary between the QBT semimetal and the QAH state as shown in Fig.~\ref{fig:phase}.
We note that the relative sign between the two terms in \eq{eq:soln-eff-S} is crucial for the existence of a physical solution for the QAH order.
Further a BCS-like solution is dependent on the presence of a term proportional to $J \ln J$, and its absence eliminates the possibility of realizing a symmetry broken state at arbitrarily weak coupling as we show in Sec. \ref{sec:nematic}.

\subsection{PEGP analysis of the lattice model: QAH solution}
The effective action based derivation of the QAH order is subject to the approximations inherent in the derivation of an effective theory.
These approximations prevent a direct comparison with results obtained in numerical simulations with the lattice Hamiltonian.
In this section we work directly with the lattice model, and obtain  various properties of the phase diagram, some of which deviate both qualitatively and quantitatively from those obtained in Section \ref{sec:eff-S}.
First we contrast the behavior of the QAH order on the $V_1$ and $V_2$ axes.
Next we determine the region in the two dimensional phase diagram where a QAH state is present, and argue for the qualitative accuracy of the phase boundary obtained in Section \ref{sec:eff-S}.

\subsubsection{$V_1 > 0, V_2 = 0$}

\begin{figure}[!t]
\centering
\includegraphics[width=1\linewidth]{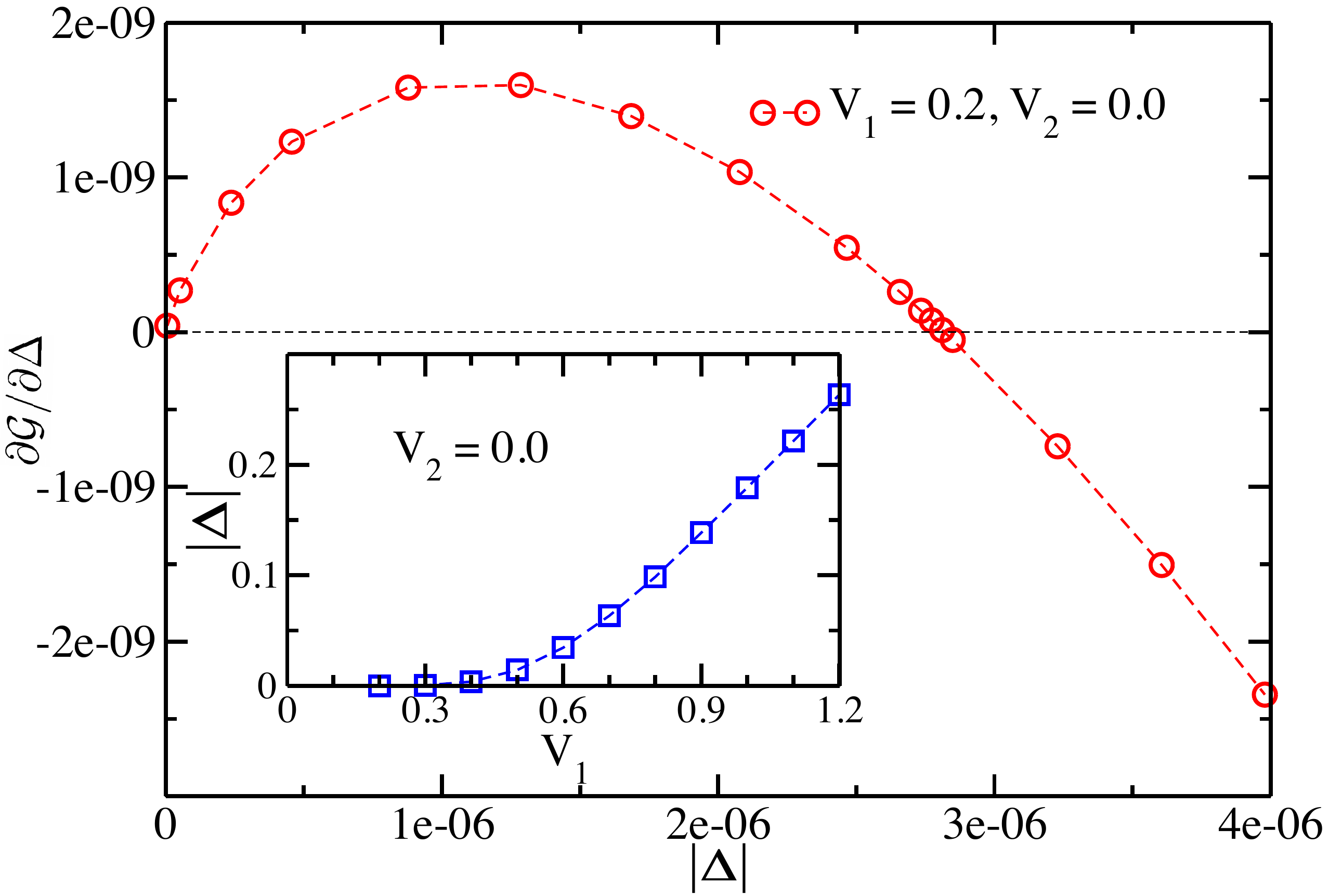}
\caption{The PEGP calculation of the QAH order for the $V_1$ model. The main figure shows the $\Delta$ dependence of $\partial \mathcal{G} / \partial \Delta$ for $V_1 = 0.2, V_2 = 0$ in the PEGP calculation up to the first-order expansion of $V_1$. $\partial \mathcal{G} / \partial \Delta$ vanishes for  some $J = J_*>0$, indicating an extremum of the free energy. Since $\dow^2\mc{G}/\dow \Dl^2 > 0$ at $J_*$, it is in fact a minimum and corresponds to the QAH state. The inset shows the $V_1$ dependence of the QAH order parameter $\Delta$.
}
\label{fig:v1_order}
\end{figure}

On the $V_1$ axis where $V_2 = 0$, Sun \emph{et al.} \cite{sun2009} obtained the  mean-field phase diagram. 
We start by reproducing this result using the PEGP method up to a first-order expansion of the free energy in $V_1$. 
The details are provided in Appendix \ref{app:pegp-qah}.

For any given $V_1 > 0$, $\partial \mathcal{G} / \partial \Delta = 0$ has a solution in terms of $J>0$ which leads to a solution for the QAH order parameter, $\Dl$. 
It suggests that any small repulsive $V_1$  would drive a QAH phase. 
In \fig{fig:v1_order} we demonstrate a representative behavior of $\partial \mathcal{G} / \partial \Delta$ as a function of $\Dl$.
In the inset of Fig.~\ref{fig:v1_order} we show the $V_1$ dependence of $\Dl$ obtained  from the solutions above. 
At weak coupling, PEGP calculation finds $V_1 \Delta \sim \exp(-1/V_1)$, which decreases exponentially with $V_1$, and, thus, is very small in the weak interaction regime.
Both the PEGP and mean-field results~\cite{sun2009} indicate that it would be extremely hard to identify the QAH phase in the weak interaction regime by numerical simulation because of the very large correlation length. 
Only in the intermediate regime where the gap becomes large enough, the order would be potentially detectable in numerical simulations.

\subsubsection{$V_1 = 0, V_2 \neq 0$}
\begin{figure}[!t]
\includegraphics[width=1\linewidth]{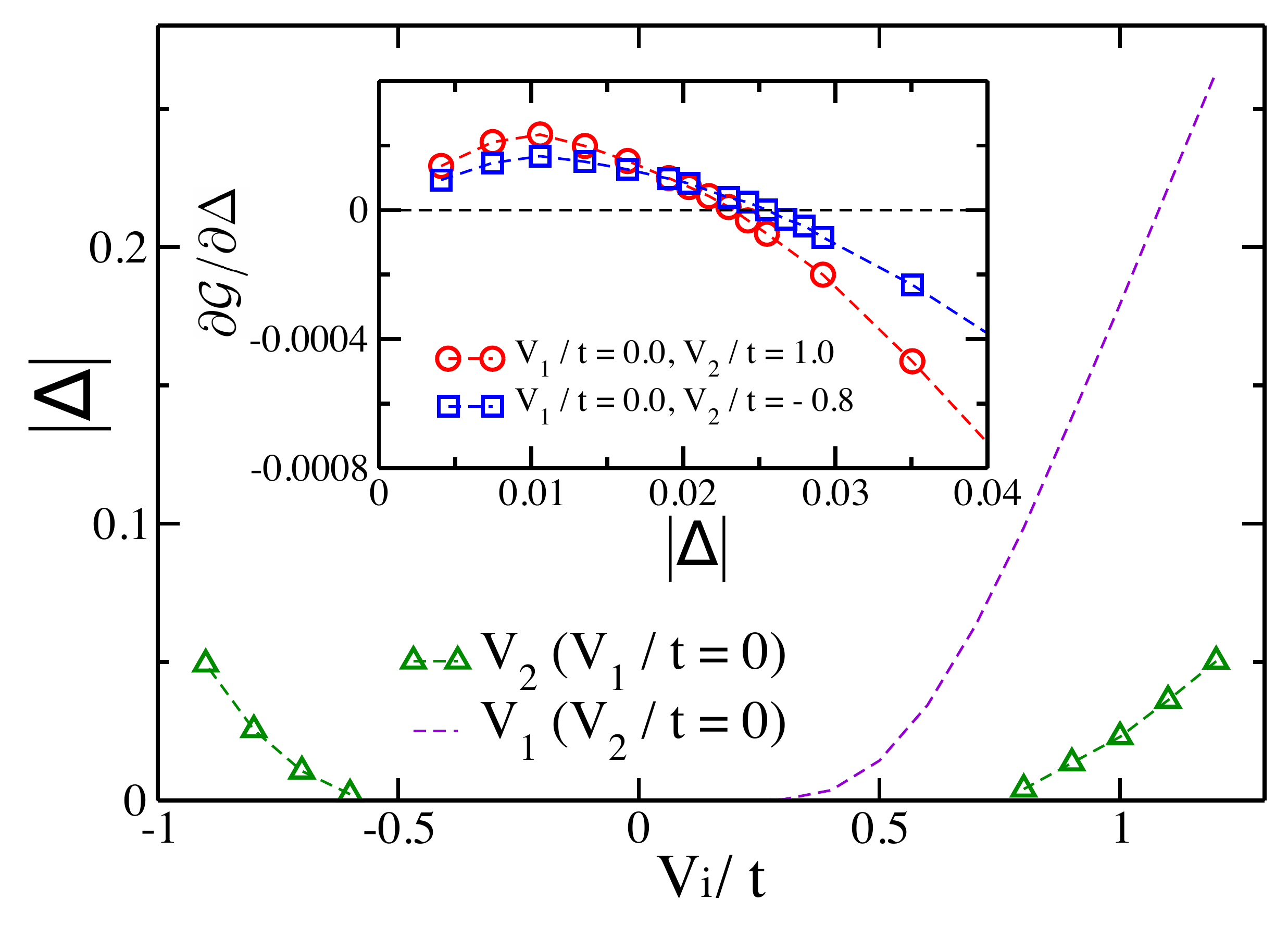}
\caption{The PEGP calculation of the QAH order for the $V_2$ model.
The main figure shows the $V_2$ dependence of $\Delta$ obtained from the PEGP calculation up to the second-order expansion of $V_2$ interaction. For comparison, we also show the $V_1$ dependence of $\Delta$ by  the dashed line, which is from the inset of Fig.~\ref{fig:v1_order}. The inset shows the $\Delta$ dependence of $\partial \mathcal{G} / \partial \Delta$ for $V_1 = 0, V_2 = 1.0, -0.8$ in the PEGP calculation up to the second-order expansion of $V_2$. 
}
\label{fig:v2_order}
\end{figure}

In this subsection, we study the model in Eq.~\eqref{eq:h} with only $V_2$ interaction, which is new to the best of our knowledge.
In the low energy effective theory the $V_2$ vertex leads to derivative coupling which makes it irrelevant in an RG sense.
Therefore, it is not directly  considered in the presence of the $V_1$  interaction vertex which leads to a marginal operator.
The magnitude of $V_2$, however, affects the energy scales in the symmetry broken states because the bare value of effective marginal coupling depends on both $V_1$ and $V_2$ as demonstrated in Section \ref{sec:eff-S}.
A crucial advantage of the PEGP over conventional mean-field strategies is apparent in this analysis, since the $V_2$ term in \eq{eq:ham} cannot be easily transformed into a mean-field theory of the QAH ordered state.
The PEGP, being independent of an a priori choice of the symmetry broken state, can be applied in analogy to the $V_1$-only model.

In the context of the PEGP calculations the key difference between the $V_1$-only and $V_2$-only models appears in the absence of the $J \ln J$ term at linear order in the latter.
Owing to the absence of the $J \ln J$ term, $\partial \mathcal{G} / \partial \Delta = 0$ does not have a non-trivial solution at arbitrary  $V_2$ which is in contrast to the  presence of a solution for any $V_1>0$.
A non-trivial solution, however, appears at quadratic order in $V_2$, reflecting the fact that quantum fluctuations of the $V_2$ vertex generates an effective marginally relevant  vertex.
Since the solution appears at order $V_2^2$, its existence is independent of the sign of $V_2$, albeit its precise value is sensitive to the sign of $V_2$ through the linear-$V_2$ term in the expression of the free energy.
The linear-$V_2$ term produces an asymmetry of the QAH order along the $V_2$ axis as seen in \fig{fig:v2_order}, where  we plot the $V_2$ dependence of $\Delta$, and show that both repulsive and attractive $V_2$ lead to a QAH state.  
This asymmetric dependence on $V_2$ is missed by the analysis in Section \ref{sec:eff-S}.
By comparing the $V_2$ dependence of $\Dl$ with the $V_1$ dependence  in \fig{fig:v1_order}, we note that the QAH order driven by $V_2$ is much weaker than that driven by $V_1$.

\subsubsection{$V_1, V_2 \neq 0$}

\begin{figure}[!t]
\includegraphics[width=1\linewidth]{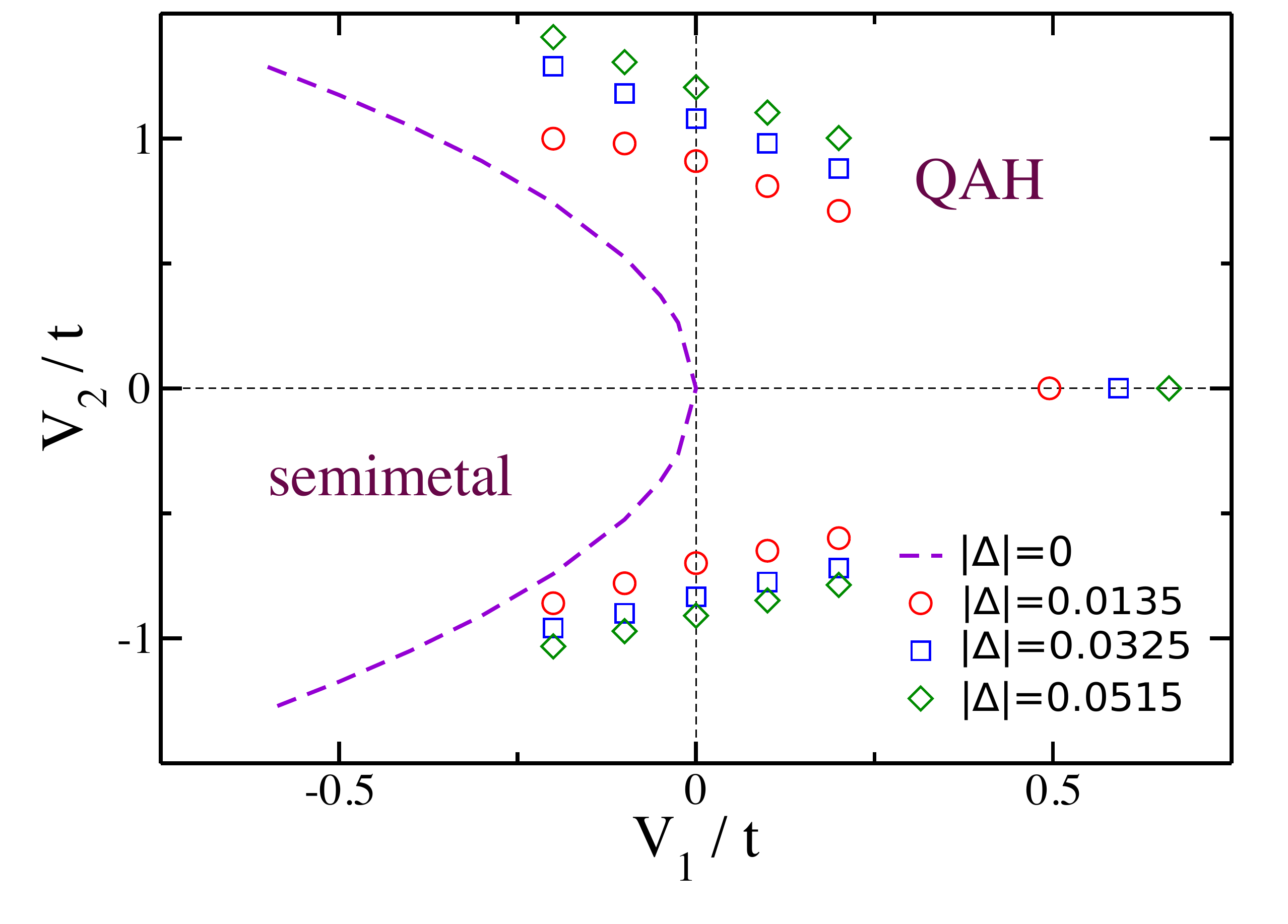}
\caption{Contour plot of the QAH order on the $V_1 - V_2$ plane.
The dashed line $4 V_1 + 1.45 V_2^2 = 0$ denotes the phase boundary between the semi-metal
and the QAH phase, which is determined from the effective low energy theory. The symbol data
are obtained from the PEGP calculation up to  
second-order in $V_2$.
}
\label{fig:qah_order}
\end{figure}

As shown in Section \ref{sec:eff-S}, the bare value of the effective coupling, $g_0$, is set by the lattice interaction strengths, $V_1$ and $V_2$.
In general $g_0$ can change sign depending on the sign and magnitude of $V_1$ and $V_2$.
Indeed, the weak-coupling expression of $g_0$ suggests that the effective coupling is attractive for a sufficiently attractive $V_1$.
RG analysis, however, imply that for an attractive $g_0$ interactions are marginally irrelevant and the QBT semi-metal is stable at weak coupling.
Therefore, we expect that in the region of the phase diagram where $V_1<0$ there exists a phase boundary separating the semi-metal from the QAH phase.
An asymptotic expression of the phase boundary was derived in Section \ref{sec:eff-S}.
Here we utilize the lattice model and argue that a phase boundary is indeed present on the $V_1<0$ half-plane, and it  qualitatively resembles the one deduced from the effective theory.

The PEGP based analyses suggest that both $V_1$ (repulsive) and $V_2$ (repulsive and attractive) interactions can independently drive the semi-metal into a QAH state.
We repeat the same calculation in the presence of both $V_1$ and $V_2$.
For simplicity we focus on the region where $|V_1| \sim |V_2|^2 \lesssim 1$, such that up to quadratic order in the expansion of the free energy we ignore terms on the order of $V_1 V_2$.
Since the QAH instability is driven by a marginally relevant interaction, the QAH gap decays exponentially on approaching the phase boundary which makes it difficult to numerically access the region around the boundary.
Nevertheless, it is still possible to identify qualitative features of the phase boundary by mapping out contours of constant magnitude of  $\Dl$ as shown in \fig{fig:qah_order}.
We note that as $\Dl$ decreases the contours approach the asymptotic phase boundary.


\section{Absence of a nematic state at weak coupling} \label{sec:nematic}
In Ref. \cite{sun2009} Sun \emph{et al.} showed that the runaway flow of $g$ in \eq{eq:eff-S} due to quantum fluctuations potentially leads to three distinct states, site and bond nematic orders, and the QAH. 
From a mean-field analysis the dominance of the QAH state was established in the absence of $V_2$ with $V_1>0$.
In the presence of an attractive $V_2$, however, the authors argued that a nematic semi-metallic state is dominant for sufficiently large $|V_2|/V_1$. 
In this section we show that such a nematic semimetal is in fact subdominant to the fully gapped QAH state through (i) an explicit susceptibility analysis within the effective field theory in \eq{eq:eff-S}, (ii) a PEGP based analysis of the lattice model, and (iii) finite-size scaling behavior of DMRG results.
\subsection{Susceptibility analysis and PEGP calculation}
In order to compare the susceptibilities of potential symmetry broken states, we start with the effective model where modes carrying momenta above an emergent scale ${\Lam_0}$ have been integrated out, and all irrelevant terms are dropped. 
As shown in Appendix \ref{app:suscep} the interaction strength flows as \cite{sun2009},
\begin{align}
g(\ell) = \frac{g(0)}{1- \frac{\ell}{\ell_c}},
\label{eq:gl}
\end{align}
where $\ell \equiv \ln{(\Lam_0/\Lam)}$ with $\Lam_0 > \Lam$ is the RG distance, and $\ell_c \equiv \frac{2\pi}{g(0)}$.
A repulsive $g(\ell)$ flows to strong coupling as $\ell$ approaches $\ell_c$ from below.
We introduce test vertices, $-\Dl_j \int dr \Psi^\dag(r) \sig_j \Psi(r)$ where $j=1,2,3$, and obtain the evolution of the source,  $\Dl_j(\ell)$, under RG flow in units of $\Dl_2(\ell)$,
\begin{align}
\frac{\Dl_j(\ell)}{\Dl_2(\ell)} =  \lt(1 - \frac{\ell}{\ell_c} \rt)^{A_2 - A_j} \frac{\Dl_j(0)}{\Dl_2(0)},
\end{align}
where $2A_1 = 2A_3 = A_2 = 1$.
Therefore, as the system flows to a strongly interacting theory $\Dl_j/\Dl_2$ with $j=1,3$ vanishes, indicating a dominant tendency for condensation of the QAH order parameter, $\Psi^\dag(r) \sig_2 \Psi(r)$ as shown in Appendix \ref{app:suscep}.
An explicit computation of the evolution of the respective susceptibilities confirms this  expectation.
In particular, as $\ell \rtarw \ell_c$ the QAH susceptibility diverges algebraically, $\chi_2(\ell) \sim (\ell_c - \ell)^{-1}$, while the nematic susceptibilities diverge logarithmically, $\chi_j(\ell) \sim \ln(\ell_c - \ell)$.
It is interesting that all susceptibilities diverge, albeit with varying rates.
We note that, although our choice of the hopping parameters enhances the symmetry of the non-interacting part of the effective action in \eq{eq:eff-S} as shown in Appendix \ref{app:suscep}, the QAH state remains the dominant instability even in the absence of the symmetry.

We arrive at the same conclusion from an explicit computation of the site-nematic order from the lattice theory with the help of the PEGP method.
To simplify the analysis we set $V_1 = 0$ in the lattice model which  realizes the extreme limit of $|V_2|/V_1 \rtarw \infty$.
We focus on the site-nematic ordering and introduce the source, $J_{nem} \int dk \Psi^{\dag}(k) \sig_3 \Psi(k)$ to the action in \eq{eq:S}.
The propagator in the presence of the source is given by,
\begin{align}
G(k; J_{nem}) = \frac{ik_0 + d_1(\bs k) \sig_1 + (J_{nem} + d_3(\bs k))\sig_3 }{k_0^2 + d_1^2(\bs k) + (J_{nem} + d_3(\bs k))^2 },
\end{align}
and the site nematic order is 
\begin{align}
\Dl_{nem}(J_{nem}) = -\int \dd{k} \tr{\sig_3 G(k; J_{nem})}.
\end{align}
As derived in Appendix \ref{app:pegp-nem}, at linear order in $V_2$ the Gibbs free energy takes the form,
\begin{align}
\mc{G}(\Dl_{nem}) &=  \mc{G}_0(\Dl_{nem}) \nn \\
& + (2\pi)^3 \dl^{(3)}(0) V_2 \lt[ 2\Dl_{nem}^2 - I_x^2 - I_y^2 \rt],
\label{eq:G-nem}
\end{align}
where $I_\mu(J) = \int \dd{\bs k} \cos(k_\mu) \frac{J + d_3(\bs k)}{M(\bs k; J)}$ with $M(\bs k; J) = \sqrt{d_1^2(\bs k) + (J + d_3(\bs k))^2}$.
The most singular term [proportional to $(J\ln{J})^2$] in the sum  $I_x^2(J_{nem}) + I_y^2(J_{nem})$ exactly cancels the singular term resulting from $2 \Dl_{nem}^2(J_{nem})$, which implies an absence of a non-trivial solution of $\dow_{\Dl_{nem}} \mc{G}(\Dl_{nem}) =0$ for arbitrary $V_2$.
Therefore, the site-nematic order is absent at small $V_2$ with $V_1 = 0$.

While results from both methods discussed above agree, they are most robust as long as the interactions are weak.
In the following we support the conclusion by large-scale DMRG calculations.

\subsection{DMRG results}
\begin{figure}[!t]
\includegraphics[width=1\linewidth]{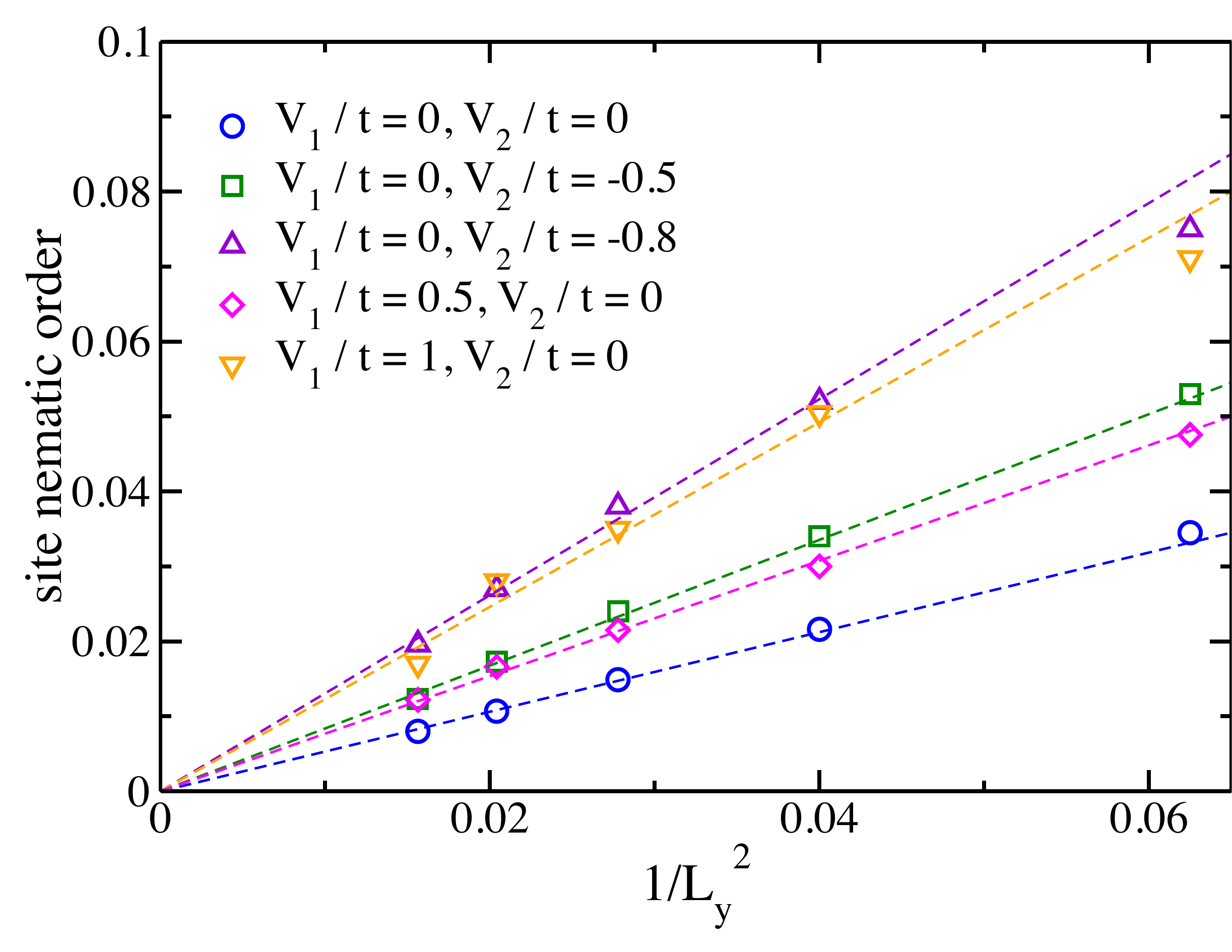}
\caption{Finite-size scaling of the site nematic order versus $1/L_y^2$. The cylinder system has  even $L_y$ with the periodic boundary conditions and odd $L_y$ with the anti-periodic boundary conditions for $L_y = 4, 5, 6, 7, 8$. The model has either $V_1 > 0$ or $V_2 < 0$. All the DMRG data are fitted linearly to $1/L_y^2$.
}
\label{fig:nematic}
\end{figure}
In our DMRG calculation of the site nematic order, we use the cylinder geometry as shown in Fig.~\ref{fig:model}(a).
Since the mirror symmetry between the two sublattices is broken on the cylinder, the site nematic order $\Dl_{nem}$ would be nonzero for  finite $L_y$. 
If the site-nematic metal phase exists, $\Dl_{nem}$ should be finite in the thermodynamic limit; otherwise, it would scale to zero with growing $L_y$. 
Here, we numerically calculate the site-nematic order $\Dl_{nem}$ on the cylinder  with  even $L_y$ for the periodic boundary conditions, and odd $L_y$ for the anti-periodic boundary conditions. 
In the non-interacting limit $\Dl_{nem}$ is expected to vanish as $L_y \rtarw \infty$.
We find that the DMRG data in this limit scale as $\Dl_{nem} \sim 1/L_y^2$ (shown in Fig.~\ref{fig:nematic}).
For weak $V_1 > 0$ or $V_2 < 0$,  $\Dl_{nem}$ also seems to scale to zero as $1/L_y^2$, indicating an absence of the site-nematic order which is consistent with our analytical results. 
We extend the DMRG calculation to the region near the phase boundary and consider the points $V_1 = 1.0$ and  $V_2 = -0.8$ on the $V_1$ and $V_2$ axis, respectively. 
The data show small oscillations, which may be attributed to strong fluctuations near the phase boundary. 
Overall, the data seem to still follow the $1/L_y^2$ scaling behavior and extrapolate to zero.
Although the nematic metal  phase is absent at weak coupling in the present model, analogous phases may be stabilized in the absence of time reversal symmetry.
Indeed in Ref. \cite{balazs2014} the authors show that weak interactions can drive a QBT semimetal that breaks time-reversal symmetry but not the rotational  symmetry  in to a nematic semimetal  state within a suitable range of hopping parameters. 


\section{Conclusion and discussion}
In this work we studied a system of spinless fermions on the checkerboard lattice in the presence of competing interactions.
In the non-interacting limit a quadratic band touching (QBT) semimetal is realized at half-filling.
The semimetallic state is protected by time-reversal and fourfold rotational symmetries.
Spontaneously breaking these symmetries leads to various symmetry broken states in the presence of interactions. 
We used a combination of numerical  (density matrix renormalization group or DMRG) and analytic (power expanded Gibbs potential or PEGP, and renormalization group) methods to obtain the quantum phase diagram of the system at half filling in \fig{fig:phase}.
The PEGP method is expected to serve as an alternative to mean-field theory when the latter is unambiguously applicable, and enables a systematic accounting for higher order corrections to mean-field based results.
Moreover, when the formulation of a mean field description is ambiguous, the PEGP provides a clear way for  accessing the relevant physics as demonstrated in this work.

In DMRG calculation, we established a quantum anomalous Hall (QAH) phase near the region with $V_1 \sim V_2^2 \sim 4$ by compelling numerical evidence, including spontaneous time-reversal symmetry breaking and quantized topological Chern number $C = 1$.
In the weak interaction region, we utilized the PEGP method which treats $V_1$ and $V_2$ on equal footing to show  that $V_2$ interaction can also drive a QAH instability.
We identified the phase boundary that separates the QBT semimetal from the QAH state, as shown by the dashed line with $V_1 \sim - V_2^2$ in \fig{fig:phase}.
In the region with attractive $V_2$ interaction  and $|V_2| \gg V_1$, our analytic calculation and DMRG simulation do not find a nematic semimetal phase at weak coupling, which differs from  Ref. \cite{sun2009}.
Our PEGP and susceptibility analyses indicate that the QAH state is the only instability of the quadratic band touching semimetal in the presence of further-neighbor interaction.
Under the assumption of a  single-parameter scaling of correlation functions as exemplified by \eq{eq:gl} the QAH phase   obtained at intermediate-coupling and small system size must be  smoothly connected to that obtained at weaker couplings and larger system sizes. 
Therefore, the ground state of the system in the entire region to the right of the asymptotic phase boundary, enclosed by the classical phases, is QAH.

In Ref.~\cite{sun2009}, it has been pointed out that the spinful version of this model may also realize a spin triplet quantum spin Hall phase depending on the strengths of the on-site Hubbard repulsion, the nearest-neighbor repulsion and exchange interaction. 
This quantum spin Hall phase, however, has not been identified in large-scale numerical simulation, and deserves further study.
\\

Note added. While finalizing  this work we became aware of a related work \cite{zeng2018}, where the authors study the $V_1$-only model on the checkerboard lattice using DMRG.


\begin{acknowledgements}
S.S.G. thanks W. Zhu, T. S. Zeng, and D. N. Sheng for extensive discussions. 
This work was performed at the National High Magnetic Field Laboratory, which is supported by National Science Foundation Cooperative Agreements No. DMR-1157490 and No. DMR-1644779, and the State of Florida. 
S.S. and K. Y. were supported by the National Science Foundation No. DMR-1442366.
S.S. also acknowledges the hospitality of the Aspen Center for Physics, which is supported by
National Science Foundation Grant No. PHY-1607611.
O. V. was supported by NSF DMR-1506756.
S.S.G. also acknowledges the computation support of project PHY-160036 from the XSEDE and the start-up funding support from Beihang University.
\end{acknowledgements}


\newpage
\onecolumngrid
\appendix

\section{Charge density wave orders and phase transitions}
\label{app:cdw}

\begin{figure}[t]
\includegraphics[width=0.8\linewidth]{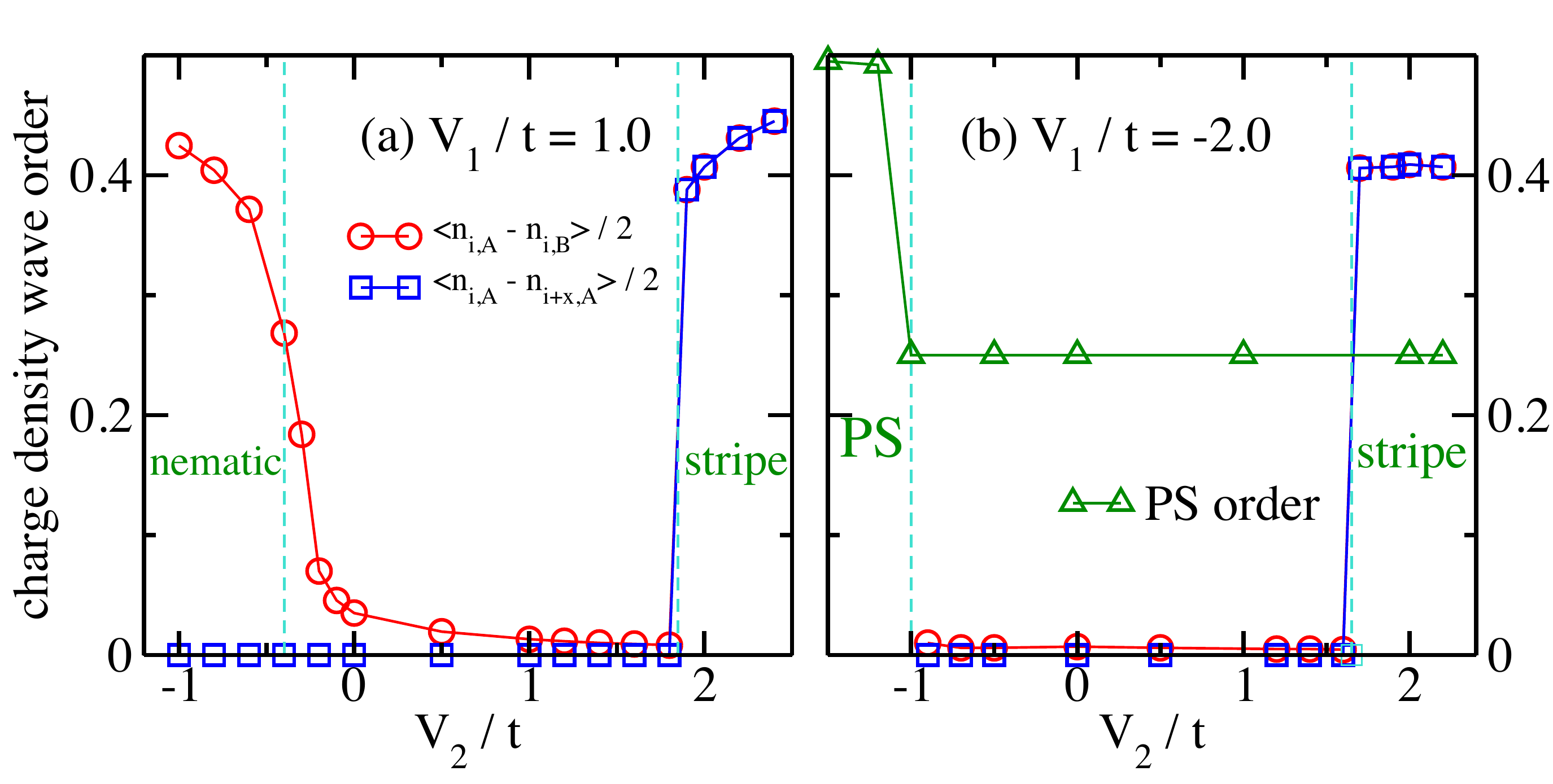}
\caption{$V_2$ dependence of the charge density wave order parameters.
(a) and (b) are for $V_1/t = 1.0$ and $-2.0$ on the $L_y = 6$ cylinder.
$\langle n_{i,A}\rangle$ and $\langle n_{i,B}\rangle$ denote the charge density of the $A$- and $B$-sublattice site in the unit cell $i$.
$\langle n_{i+\hat{x},A}\rangle$ is the density of the $A$-sublattice in the $i+\hat{x}$ unit cell.
The site nematic order and the stripe order can be characterized by $\langle n_{i,A} - n_{i,B}\rangle / 2$ and $\langle n_{i,A} - n_{i+\hat{x},A}\rangle / 2$.
The phase separation (PS) order parameter is defined as $\sum_{i \in \rm half}\langle n_i \rangle / N$, and here we show the results of the occupied half side in the PS phase.
In the site nematic and stripe insulator phase, the PS order parameter is always $1/4$.
The data in (a) and (b) have the same symbol definitions.
}
\label{fig:cdw}
\end{figure}

\begin{figure}[t]
\includegraphics[width=0.45\linewidth]{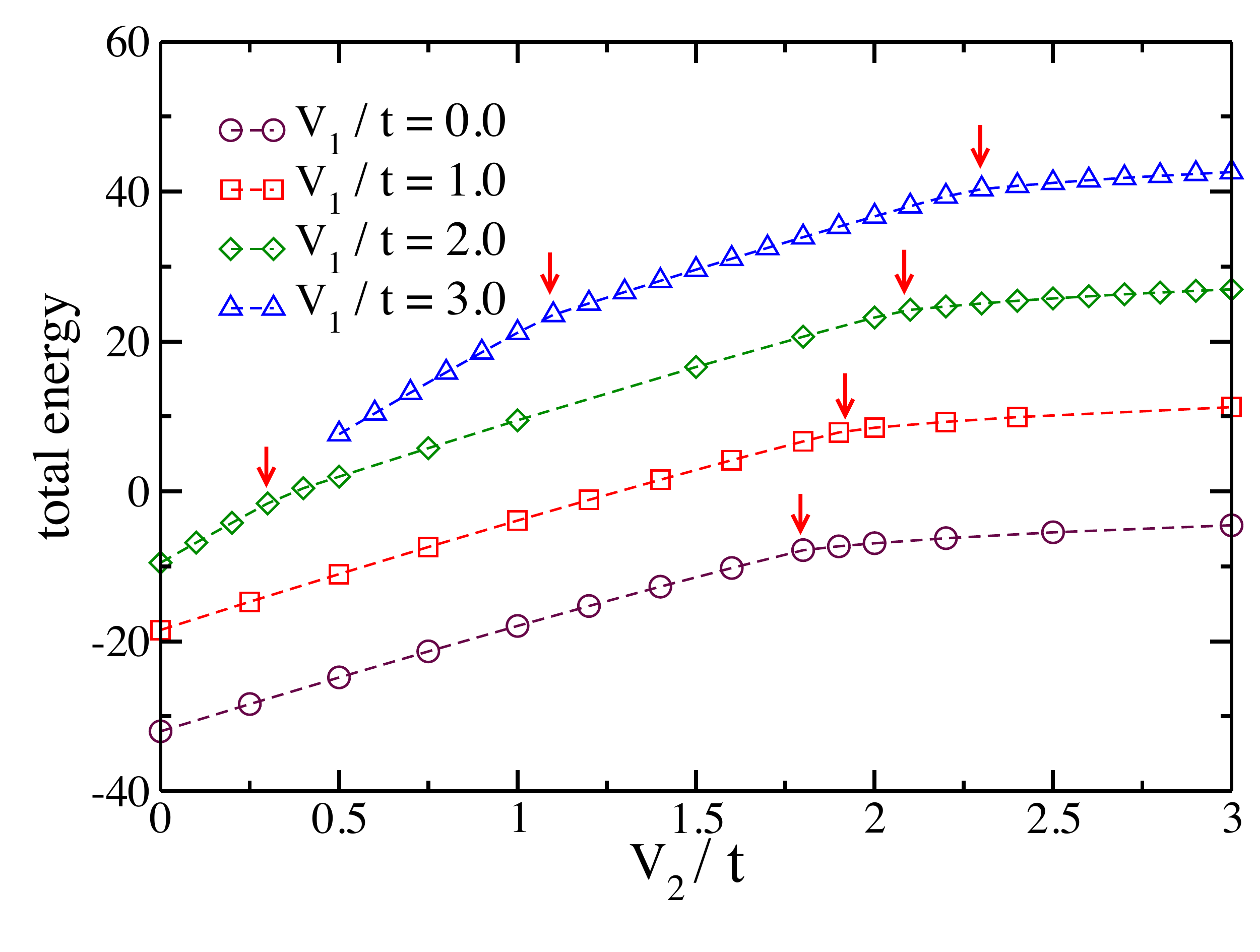}
\includegraphics[width=0.45\linewidth]{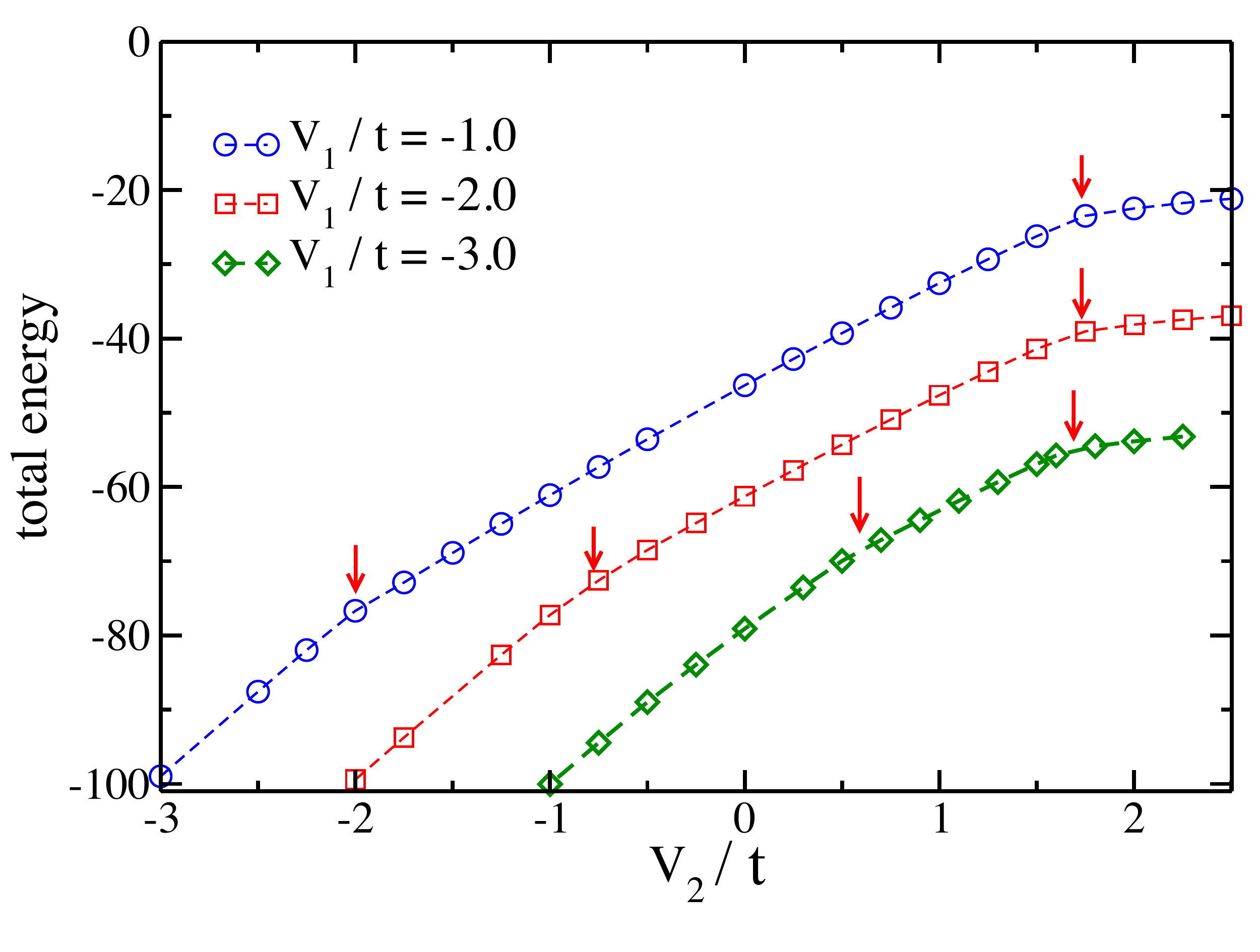}
\caption{Characterizing the quantum phase transitions to the insulator phases through ground-state energy. The total energy is obtained on the $L_x = L_y = 4$ torus system. The red arrow indicates the kink of energy that characterizes the first-order quantum phase transitions to the insulator phases as shown in Fig.~\ref{fig:phase} of the main text.
}
\label{appfig:torus}
\end{figure}

We first show the charge density wave ordered phases and the phase transitions in Fig.~\ref{fig:phase}.
As the open boundary conditions of the cylinder geometry, DMRG calculation obtains non-uniform distribution of the charge density in the charge density wave phases, which are shown in the inset of Fig.~\ref{fig:model}(b).
To characterize the charge density wave phases, we can measure three order parameters.
The first order parameter is defined as the charge density difference of the two sublattices $\langle n_{i,A} - n_{i,B}\rangle / 2$, where $\langle n_{i,A}\rangle$ ($\langle n_{i,B}\rangle$) denotes the charge density of the $A$($B$)-sublattice site in the unit cell $i$.
The second order parameter is the charge density difference of the neighboring sites in the same sublattice, i.e., $\langle n_{i,A} - n_{i+\hat{x},A}\rangle / 2$ or $\langle n_{i,B} - n_{i+\hat{x},B}\rangle / 2$.
In the site nematic phase, $\langle n_{i,A} - n_{i,B}\rangle / 2$ is finite and $\langle n_{i,A} - n_{i+\hat{x},A}\rangle / 2$ is zero; in the stripe phase, both order parameters have the same finite value.
In the phase separation region, DMRG calculation obtains the state with charges staying on either left or right side of the lattice, leaving the other half sites empty.
We can define the third order parameter as the average density of the half sites $\sum_{i \in \rm half}\langle n_i \rangle / N$, which is either $1/2$ or $0$ in the phase separation.
In the site nematic and stripe insulator phase, the phase separation order parameter is always $1/4$.
In Fig.~\ref{fig:cdw}, we show the $V_2$ dependence of different charge density wave order parameters, which show a sharp enhancement in the phase boundaries, characterizing the phase transitions.
We also show the phase transitions by studying the ground-state energy on the $L_x = L_y = 4$ torus system.
The total energy of the torus is shown in Fig.~\ref{appfig:torus}, where the energy exhibits a kink at the transition point, which suggests the first-order transitions and are consistent with the order parameter change in Fig.~\ref{fig:cdw}.

\section{PEGP calculation for QAH order} \label{app:pegp-qah}
\begin{figure}[!t]
\centering
\begin{subfigure}[b]{0.33\columnwidth}
\includegraphics[width=0.55\columnwidth]{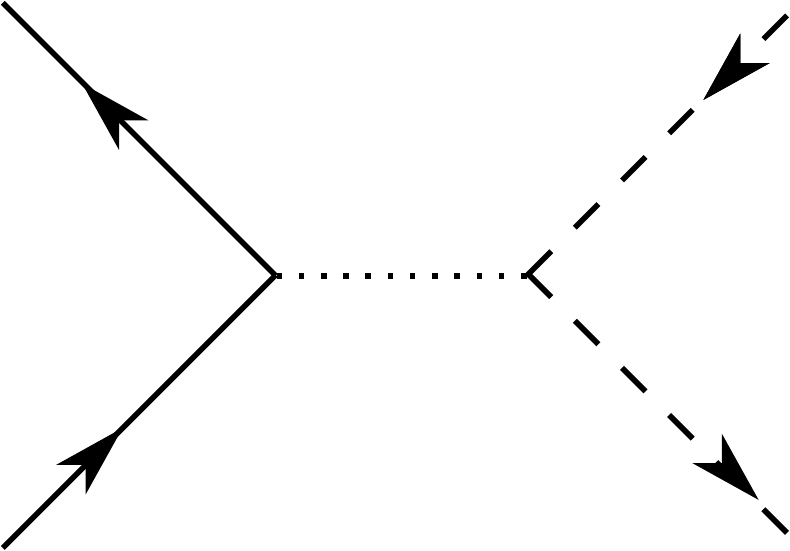}
\caption{$V_1$ vertex}
\end{subfigure}
\hfill
\begin{subfigure}[b]{0.66\columnwidth}
\includegraphics[width=0.7\columnwidth]{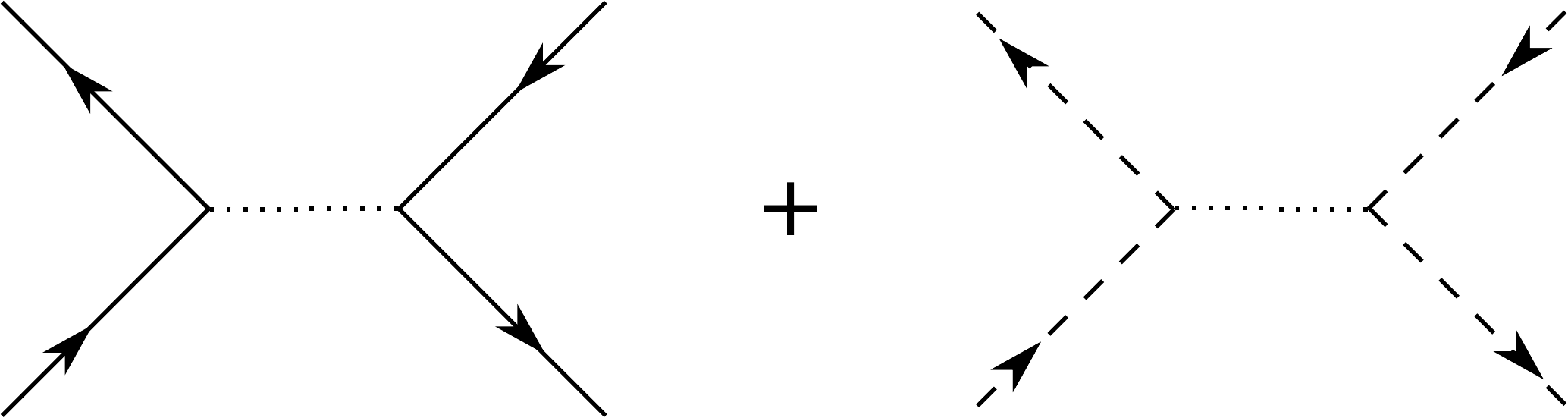}
\caption{$V_2$ vertices}
\end{subfigure}
\caption{Graphical representation of the $V_1$ and $V_2$ vertices. The solid (dashed) line represents the $\mf a$ ($\mf b$) type fermions, while the dotted line represents the momentum dependent coupling functions.}
\label{fig:S1}
\end{figure}

In this section we collect the vacuum diagrams required for the calculation of the QAH gap within the PEGP formalism.
For the calculation it is convenient to define
\begin{align}
& d_1(\bs k) = 4 \cos{\frac{k_x}{2}} \cos{\frac{k_y}{2}}; \nn \\ 
& d_2(\bs k) = 4 \sin{\frac{k_x}{2}} \sin{\frac{k_y}{2}}; \nn \\ 
& d_3(\bs k) =  \cos{k_x} -  \cos{k_y}; \nn \\
& m(\bs k,J) = \sqrt{d_1^2(\bs k) + J^2 d_2^2(\bs k) + d_3^2(\bs k)} 
= \sqrt{(2 + \cos{k_x} + \cos{k_y})^2 + \lt( 4 J \sin{\frac{k_x}{2}}  \sin{\frac{k_y}{2}} \rt)^2},
\label{eq:defn}   
\end{align}
and the operator
\begin{align}
\mc{O} = \int \dd{k} \psi^{\dag}(k) d_2(\bs k) \sig_2 \psi(k),
\end{align}
where $\psi(k) = \trans{(\mf{a}(k), \mf{b}(k))}$.
The total action is
\begin{align}
S[J] = S_0 + S_1 + J \mc{O},
\end{align}
where
\begin{align}
 S_0 &= \int \dd{k} \psi^{\dag}(k) \lt[ -ik_0 \sig_0 + d_1(\bs k) \sig_1 + d_3(\bs k) \sig_3 \rt] \psi(k), \\
 S_1 &= \int \dd{k} \dd{k'} \dd{q} \Bigl[ 4 V_1 \cos{\frac{q_x}{2}} \cos{\frac{q_y}{2}} ~ \A^{\dag}(k + q) \A(k) \B^{\dag}(k') \B(k' +  q)  \nn \\ 
& \quad + 2V_2 (\cos{q_x} + \cos{q_y} - 2) ~ \A^{\dag}(k + q) \A(k) \A^{\dag}(k') \A(k'+q) 
+ (\A \ltrtarw \B) \Bigr]. 
\end{align}
In \fig{fig:S1} we show the representation of the two interaction vertices.

\begin{figure}[!t]
\centering
\begin{subfigure}[b]{0.9\columnwidth}
\includegraphics[width=0.8\columnwidth]{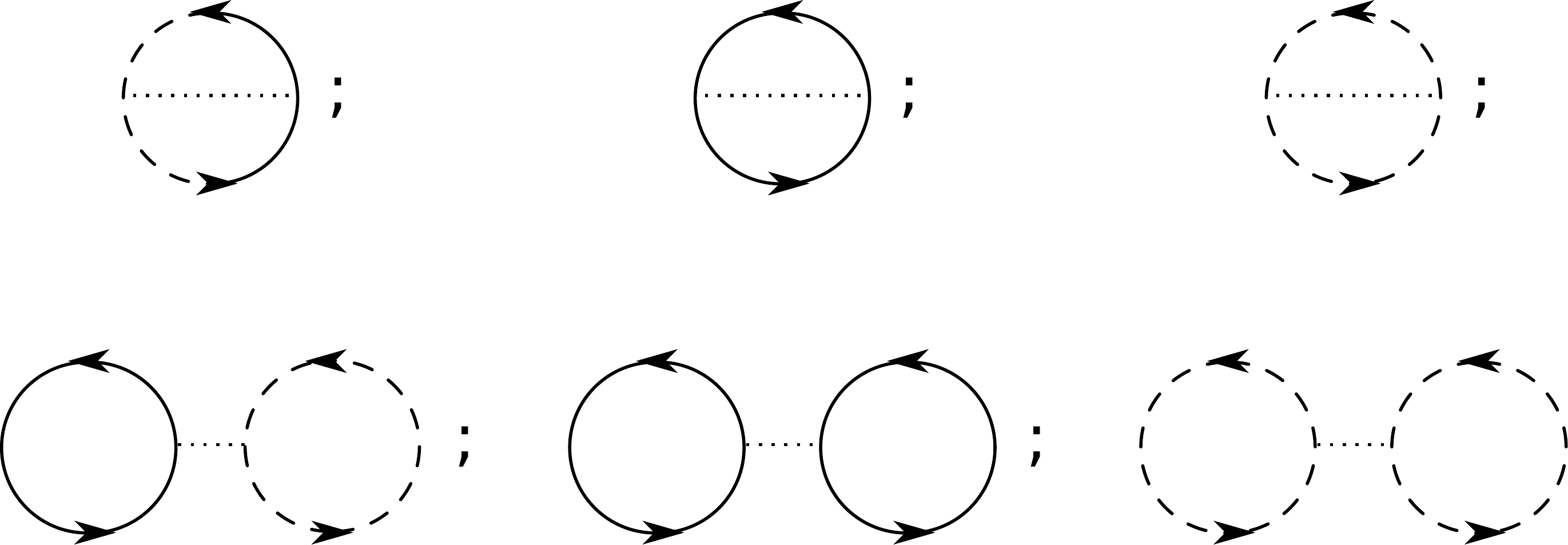}
\caption{}
\label{fig:<S1>}
\end{subfigure}
\hfill
\begin{subfigure}[b]{0.95\columnwidth}
\includegraphics[width=0.9\columnwidth]{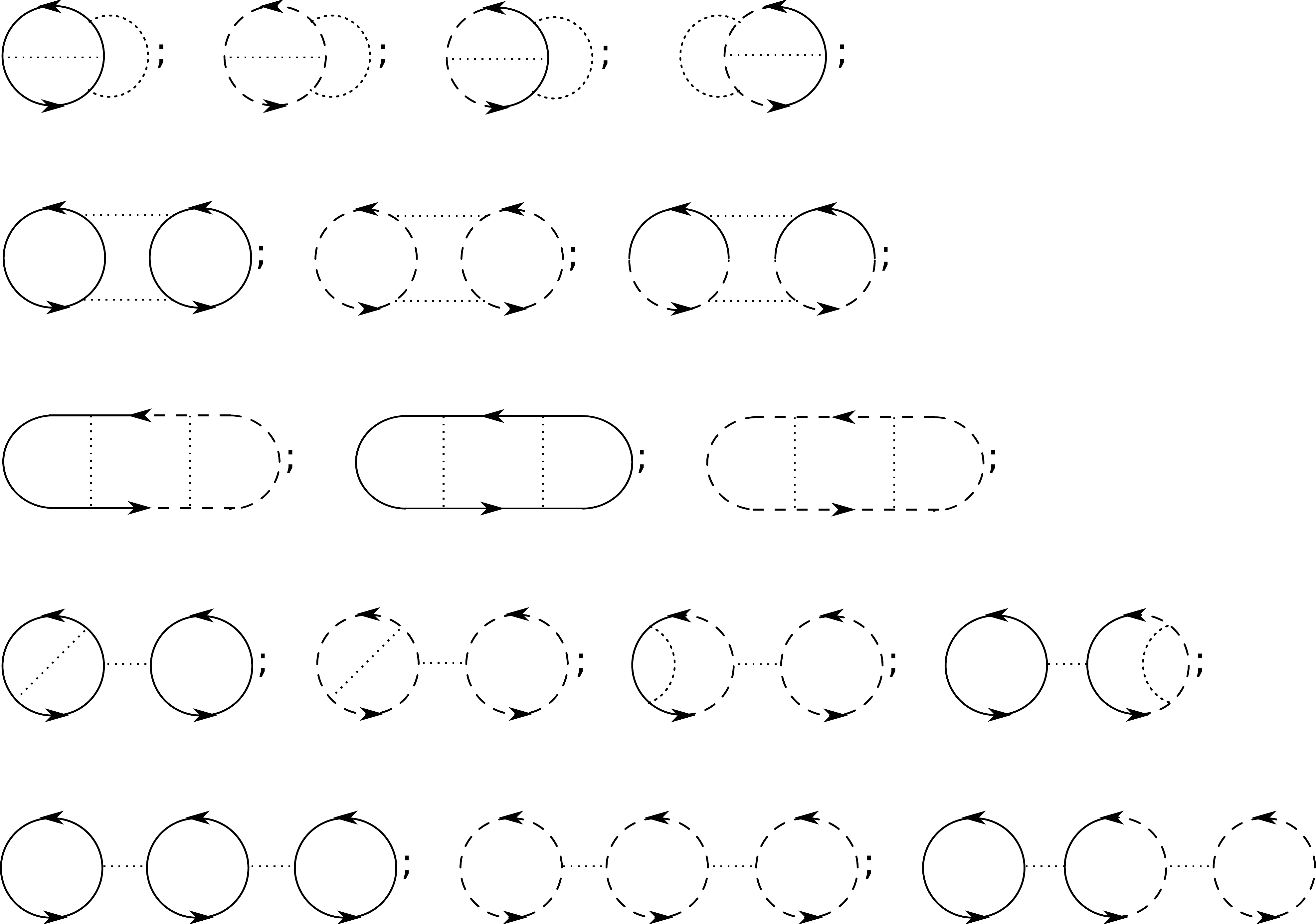}
\caption{}
\label{fig:<S1S1>}
\end{subfigure}
\caption{Vacuum diagram that contribute to $\mc{G}$.
(a) Diagrams contributing to $\avg{S_1}$. (b) Diagrams proportional to $V_2^2$ that  contribute to $\avg{S_1^2}$. The non-1PI diagrams resulting only from the $V_2$ vertex vanishes identically.}
\label{fig:diagram}
\end{figure}

The Gibbs free energy,
\begin{align}
\mc{G}[\Dl] = - \frac{1}{\beta} \ln{\mc{Z}[J]} - L^2 J \Dl,
\end{align}
where $L^2$ is the volume of the system and $\Dl$ is the ground state expectation value of the order parameter.
Since the minima of $\mc{G}$ correspond to locally stable phases, here we are interested in those minima which occur at $\Dl \neq 0$. 
\begin{align}
\dow_{\Dl} \mc{G}(J) &= -J + T (\dow_{\Dl} J) 
\Biggl[ \dow_J \avg{S_1} \lt\{1 + \frac{T}{L^2} (\dow_J(\dow_\Dl J)) \avg{\mc{O} S_1} + \frac{T}{L^2} (\dow_{\Dl} J) (\dow_J \avg{\mc{O} S_1})  \rt\} \nn \\
&\qquad + (\dow_\Dl J)(\dow_J^2 \avg{S_1}) \avg{\mc{O} S_1} - \half \dow_{J} \avg{S_1^2} \Biggr]
\end{align}
with
\begin{align}
&\avg{S_1} = - (2\pi)^3 \dl^{(3)}(0) 
\lt[
V_1 \lt\{ f_{11}^2(J) + J^2 f_{11}^2(J) \rt\} 
+ V_2 \lt\{ f_{21}^2(J) + f_{22}^2(J)
+ f_{23a}(J) f_{23b}(J) \rt\}
\rt] \\
&\avg{S_1^2}_{V_1 = 0} = (2\pi)^3 \dl^{(3)}(0) V_2^2 \lt[ h_a(J) + h_b(J) - h_c(J) \rt] \\
& \avg{\mc{O} S_1} = (2\pi)^3 \dl^{(3)}(0) J \lt[V_1 \Om_1(J) - V_2 \Om_2(J) \rt]
\end{align}
where
\begin{align}
& f_{11}(J) = \int \dd{\bs k} \frac{d_1^2(\bs k)}{m(\bs k, J)} \\
& f_{12}(J) = \int \dd{\bs k} \frac{d_2^2(\bs k)}{m(\bs k, J)} \\
& f_{21}(J) = \int \dd{\bs k} \sin^2(k_x) \frac{d_3(\bs k)}{m(\bs k, J)} \\
& f_{22}(J) = 2 \int \dd{\bs k} \sin^2\lt(\frac{k_x}{2}\rt) \frac{d_3(\bs k)}{m(\bs k, J)} \\
& f_{23a}(J) = 2 \int \dd{\bs k} \cos{\frac{k_x}{2}} \sin^2(k_x) \frac{d_3(\bs k)}{m(\bs k, J)} \\
& f_{23b}(J) = 2 \int \dd{\bs k} \cos{\frac{k_x}{2}} \cos^2(k_x) \frac{d_3(\bs k)}{m(\bs k, J)} \\
& h_a(J) -h_c(J) = 2 \int \dd{\bs k} \dd{\mbf p} \dd{\mbf q} 
\frac{[\cos{p_x} + \cos{p_y} - 2][\cos{p_x} + \cos{p_y} - \cos{(k_x - q_x)} - \cos{(k_y - q_y)} ]}{m(\mbf{p} + \mbf{q},J) + m(\mbf{q},J) + m(\mbf{p} + \mbf{k},J ) + m(\mbf{k},J)} \nn \\
& \times \frac{1}{m(\mbf{p} +  \mbf{q},J)  m(\mbf{q},J)  m(\mbf{p} + \mbf{k},J )  m(\mbf{k},J)} \nn \\
& \times \Bigl[
\lt\{ m(\mbf q, J) m(\bs k, J) - d_3(\mbf q) d_3(\bs k) \rt\}
\lt\{ m(\mbf p + \mbf q, J) m(\mbf p + \bs k, J) - d_3(\mbf p + \mbf q) d_3(\mbf p + \bs k) \rt\} \nn \\
& \quad + 2 m(\mbf p + \bs k, J) d_3(\mbf p + \mbf q)\lt\{ m(\mbf q, J) d_3(\bs k) - d_3(\mbf q) m(\bs k, J) \rt\} 
+ d_1(\mbf p + \mbf q) d_1(\mbf q) d_1(\mbf p + \bs k) d_1(\bs k) \nn \\
& \quad + 2 J^2 d_1(\mbf p + \mbf q) d_1(\mbf q) d_2(\mbf p + \bs k) d_2(\bs k)
+ J^4 d_2(\mbf p + \mbf q) d_2(\mbf q) d_2(\mbf p + \bs k) d_2(\bs k) 
\Bigr] \\
& h_b(J) = \int \dd{\bs k} 
\lt[2 \int \dd{\mbf q} (\cos{q_x} \cos{k_x} + \cos{q_y}\cos{k_y} - 2) \frac{d_3(\mbf q)}{m(\mbf q, J)} \rt]^2 
~ \frac{d_1^2(\bs k) + J^2 d_2^2(\bs k)}{m^3(\bs k, J)} \\
& \Om_1(J) = \frac{1}{8} \int \dd{\bs k} \dd{\mbf p} 
\lt[ 
d_2^2(\bs k) d_2^2(\mbf p) 
~ \frac{d_1^2(\bs k) + d_3^2(\bs k)}{m(\mbf p, J) m^3(\bs k, J)}
- d_1^2(\bs k) d_1^2(\mbf p) 
~ \frac{d_2^2(\bs k)}{m(\mbf p, J) m^3(\bs k, J)}
\rt] \\
& \Om_2(J) = 2\int \dd{\bs k} \dd{\mbf p} 
\lt[ 
\cos{k_x} \cos{p_x} + \cos{k_y} \cos{p_y} - 2\rt]
~ \frac{d_2^2(\bs k) d_3(\bs k)}{m^3(\bs k, J)}
\frac{d_3(\mbf p)}{m(\mbf p, J)}
\end{align}
The details of the vacuum diagrams which contribute to the Gibbs free energy are demonstrated in Fig.~\ref{fig:diagram}.


\section{Susceptibility of interacting quadratic band touching: one valley and spinless} \label{app:suscep}
The susceptibilities for the QAH state and the two nematic metallic states at the non-interacting fixed point diverge, indicating a potential for realizing one or more of these states in the presence of interactions.
We compute the susceptibilities in the presence of interaction, and track their evolution under coarse graining.
We find that although all three susceptibilities tend to diverge in a finite RG time, they do so at different rates.
In particular, the susceptibility for QAH state diverges exponentially faster than the nematic states.

We start with the Hamiltonian for the effective low energy theory discussed in Section \ref{sec:eff-S},
\begin{eqnarray}
H&=&H_0+H_{int}\\
H_0&=&\sum_{|\bk|<{\Lam_0}}\Psi^\dagger_\bk\left(\tau_3\frac{k^2_x-k^2_y}{2}+\tau_1\frac{2k_xk_y}{2}\right)\Psi_\bk\\
H_{int}&=& g\int d^2\br\; \psi_a^\dagger(\br)\psi_b^\dagger(\br)\psi_b(\br)\psi_a(\br)
=\frac{1}{4}g\int d^2\br\; \Psi^\dagger(\br)\tau_2\Psi^*(\br)\Psi^T(\br)\tau_2\Psi(\br).
\end{eqnarray}
We first derive the RG flow of the coupling $g$ whereby we reproduce the result in Ref. \cite{sun2009}.
Next we derive the RG flows of the susceptibilities which are new results.

\subsection{Renormalization of the coupling}
Here we derive the RG flow of $g$.
The interaction term,
\begin{eqnarray}
S_{int}&=&\frac{1}{4}g\int d\tau\int d^2\br\; \Psi^\dagger(\br,\tau)\tau_2\Psi^*(\br,\tau)\Psi^T(\br,\tau)\tau_2\Psi(\br,\tau).
\end{eqnarray}
The quantum correction is produced by integrating out the high-energy modes \cite{shankar1994},
\begin{eqnarray}
\left\langle e^{-S_{int}}\right\rangle_> &\approx&  e^{-\left\langle S_{int}\right\rangle_>+\frac{1}{2}\left(\left\langle S^2_{int}\right\rangle_>-\left\langle S_{int}\right\rangle^2_>\right)},
\end{eqnarray}
which leads to,
\begin{eqnarray}
-\delta S_{eff}&=&\frac{1}{2}\cdot\frac{g^2}{16}\int_1\int_2
\left\langle \Psi^\dagger(1)\tau_2\Psi^*(1)\Psi^T(1)\tau_2\Psi(1)\Psi^\dagger(2)\tau_2\Psi^*(2)\Psi^T(2)\tau_2\Psi(2)\right\rangle_>\\
&=&\frac{1}{2}\cdot\frac{g^2}{16}\cdot 4\int_1\int_2 \mbox{Tr}\left[\tau_2 G_>(1-2) \tau_2 G_>^T(1-2)\right]
\Psi^\dagger(1)\tau_2\Psi^*(1)\Psi^T(2)\tau_2\Psi(2)\nonumber\\
&+&
\frac{1}{2}\cdot\frac{g^2}{16}\cdot 16\int_1\int_2 \Psi^\dagger(1)\tau_2G^T(2-1)\tau_2\Psi(2) \Psi^\dagger(2)\tau_2 G^T(1-2)\tau_2\Psi(1)\\
&\approx&\frac{1}{2}\cdot\frac{g^2}{16}\cdot 4\int\frac{d\omega}{2\pi}\int^{\Lam_0}_{\frac{{\Lam_0}}{s}} \frac{d^2\bk}{(2\pi)^2}
\mbox{Tr}\left[\tau_2 G_\bk(i\omega) \tau_2 G_{-k}^T(-i\omega)\right]
\int_1\Psi^\dagger(1)\tau_2\Psi^*(1)\Psi^T(1)\tau_2\Psi(1)\nonumber\\
&+&
\frac{1}{2}\cdot\frac{g^2}{16}\cdot 16\int\frac{d\omega}{2\pi}\int^{\Lam_0}_{\frac{{\Lam_0}}{s}} \frac{d^2\bk}{(2\pi)^2} \int_1 \Psi^\dagger(1)\tau_2G_{\bk}^T(i\omega)\tau_2\Psi(1) \Psi^\dagger(1)\tau_2 G_{\bk}^T(i\omega)\tau_2\Psi(1).
\end{eqnarray}
Here we have suppressed replaced reference to the 3-momentum, $k_n$, by $n$ for notational convenience.
Because $H_0$ is symmetric (involves only $\tau_{1,3}$), we have $G_\bk(i\omega) = G_{\bk}^T(i\omega)$, and
\begin{eqnarray}
\int_{-\infty}^{\infty}\frac{d\omega}{2\pi}\int^{\Lam_0}_{\frac{{\Lam_0}}{s}} \frac{d^2\bk}{(2\pi)^2} G_\bk(i\omega) \otimes G_{\mp k}(\mp i\omega) =
\left(\pm 1\otimes 1+\frac{1}{2}\tau_1\otimes \tau_1+\frac{1}{2}\tau_3\otimes \tau_3\right)\frac{1}{4\pi}\ln s
\end{eqnarray}
Therefore
\begin{eqnarray}
\int\frac{d\omega}{2\pi}\int^{\Lam_0}_{\frac{{\Lam_0}}{s}} \frac{d^2\bk}{(2\pi)^2}
\mbox{Tr}\left[\tau_2 G_\bk(i\omega) \tau_2 G_{-k}^T(-i\omega)\right]=0
\end{eqnarray}
and
\begin{eqnarray}
\delta S_{eff}&=&-\frac{1}{2}\cdot g^2 \frac{1}{4\pi}\ln s
\left(-\int_1 \Psi^\dagger(1)\tau_21\tau_2\Psi(1) \Psi^\dagger(1)\tau_2 1\tau_2\Psi(1)\right.\nonumber\\
&+&\left.\frac{1}{2}\int_1 \Psi^\dagger(1)\tau_2\tau_1\tau_2\Psi(1) \Psi^\dagger(1)\tau_2 \tau_1\tau_2\Psi(1)+\frac{1}{2}\int_1 \Psi^\dagger(1)\tau_2\tau_3\tau_2\Psi(1) \Psi^\dagger(1)\tau_2 \tau_3\tau_2\Psi(1)\right)\\
&=&-\frac{1}{2}\cdot g^2 \frac{1}{4\pi}\ln s\nonumber\\
&\times&\int_1\left(- \Psi^\dagger(1)\Psi(1) \Psi^\dagger(1)\Psi(1)+\frac{1}{2}\Psi^\dagger(1)\tau_1\Psi(1) \Psi^\dagger(1)\tau_1\Psi(1)+\frac{1}{2} \Psi^\dagger(1)\tau_3\Psi(1) \Psi^\dagger(1) \tau_3\Psi(1)\right)\nonumber\\
&=&\left(g^2\frac{1}{\pi}\ln s \right)\frac{1}{2}\int_1 \psi^\dagger_a \psi^\dagger_b \psi_a \psi_a
\end{eqnarray}
For $s=1+d\ell$ we have
\begin{eqnarray}
\frac{dg}{d\ell}=\frac{1}{2\pi}g^2
\label{eq:gflow}
\end{eqnarray}
This recovers the Eq.(3) in \cite{sun2009} with the replacement $g \mapsto V$:
\begin{eqnarray}
\frac{d}{d\ell}\frac{V}{|t_x|}=\frac{1}{4\pi}\left(\frac{V}{|t_x|}\right)^2
\end{eqnarray}
where we note that their $t_x$ is our $1/2$.
Solving Eq.(\ref{eq:gflow})
we find
\begin{eqnarray}
g(\ell)=\frac{1}{\frac{1}{g_0}-\frac{1}{2\pi}\ell},
\end{eqnarray}
where $g_0 = g(0)$.

\subsection{Renormalization of the symmetry breaking source terms}

We now perturb the action by adding infinitesimal symmetry breaking terms
\begin{eqnarray}
S\rightarrow S-\sum_{j=1}^3\Delta_j \int d\tau \int d^2\br \Psi^\dagger(\br,\tau)\tau_j\Psi(\br,\tau)
\end{eqnarray}
Then
\begin{eqnarray}
\langle e^{-S_{int}+\sum_{j=1}^3\Delta_j \int d\tau \int d^2\br \Psi^\dagger(\br,\tau)\tau_j\Psi(\br,\tau)}\rangle_> \rightarrow
 e^{-\langle S_{int}\sum_{j=1}^3\Delta_j \int d\tau \int d^2\br \Psi^\dagger(\br,\tau)\tau_j\Psi(\br,\tau)\rangle_>}
\end{eqnarray}
So
\begin{eqnarray}
&&-\frac{1}{4}g\sum_{j=1}^3\Delta_j\left\langle \int_1 \int_2 \Psi^\dagger(1)\tau_2\Psi^*(1)\Psi^T(1)\tau_2\Psi(1) \Psi^\dagger(2)\tau_j\Psi(2)\right\rangle_>=\nonumber\\
&&g\sum_{j=1}^3\Delta_j \int_1 \int_2 \Psi^\dagger(1)\tau_2G^T(2-1)\tau^T_jG^T(1-2) \tau_2\Psi(1)=\nonumber\\
&&g\sum_{j=1}^3\Delta_j\int\frac{d\omega}{2\pi}\int^{\Lam_0}_{\frac{{\Lam_0}}{s}} \frac{d^2\bk}{(2\pi)^2}
 \int_1 \Psi^\dagger(1)\tau_2G_{\bk}^T(i\omega)\tau^T_jG_{\bk}^T(i\omega) \tau_2\Psi(1)=\nonumber\\
&&g\frac{m}{4\pi}\ln s\sum_{j=1}^3\Delta_j
\int_1 \left(-\Psi^\dagger(1)\tau_2 1\tau^T_j1 \tau_2\Psi(1)+\frac{1}{2}\Psi^\dagger(1)\tau_2 \tau_1\tau^T_j \tau_1 \tau_2\Psi(1)+
\frac{1}{2}\Psi^\dagger(1)\tau_2 \tau_3\tau^T_j \tau_3 \tau_2\Psi(1)
\right)=\nonumber\\
&&g\frac{m}{4\pi}\ln s\sum_{j=1}^3\Delta_j
\int_1 \left(\Psi^\dagger(1)\tau_1\Psi(1)+
2\Psi^\dagger(1)\tau_2 \Psi(1)+\Psi^\dagger(1)\tau_3\Psi(1)
\right).
\end{eqnarray}

This means that
\begin{eqnarray}
\frac{d\ln \Delta_j}{d\ell}=2+A_j \frac{1}{2\pi}g, 
\quad \mbox{with} ~~ A_1=A_3=\frac{A_2}{2}=\frac{1}{2}.
\end{eqnarray}

Solving the above equation gives
\begin{eqnarray}
\ln\frac{\Delta_j(\ell)}{\Delta_j(0)}&=&2\ell+A_j \frac{1}{2\pi}\int^\ell_0 d\ell' g(\ell')=2\ell+A_j\int^\ell_0 d\ell'
\frac{1}{\frac{2\pi}{g_0}-\ell'}\nonumber\\
&=&2\ell-A_j\ln\left(1-\frac{1}{2\pi}\sig_0\ell\right).
\end{eqnarray}
or
\begin{eqnarray}
\Delta_j(\ell)&=&\frac{e^{2\ell}}{\left(1-\frac{1}{2\pi}g_0\ell\right)^{A_j}}\Delta_j(0)\nonumber\\
\end{eqnarray}

\subsection{Susceptibility}
If we sum up the contribution to the free energy from the integrated out high energy modes, we can find the correction due to the source terms.
To second order, this determines the susceptibility:
\begin{eqnarray}
\chi_j(\ell)\Delta^2_j(0)&=& C_j\int_0^\ell d\ell' e^{-4\ell'}\Delta^2_j(\ell')=C_j \Delta^2_j(0)\int_0^\ell d\ell'\frac{1}{\left(1-\frac{1}{2\pi}g_0\ell'\right)^{2A_j}}
\end{eqnarray}
Clearly, the critical value of $\ell$ is $\frac{2\pi}{g_0}$, in terms of which
\begin{eqnarray}
\chi_{1,3}(\ell)&=& C_{1,3}\int_0^\ell d\ell'\frac{1}{\left(1-\frac{1}{2\pi}g_0\ell\right)}=C_{1,3}\ell_c\ln\left(\frac{1}{1-\ell/\ell_c}\right)\\
\chi_{2}(\ell)&=& C_{2}\int_0^\ell d\ell'\frac{1}{\left(1-\frac{1}{2\pi}g_0\ell\right)^2}
=C_{2}\ell_c\frac{1}{\ell_c/\ell-1}\sim\frac{1}{(\ell_c-\ell)^{\gamma_2}}.
\end{eqnarray}

So the quantum anomalous Hall susceptibility diverges as a power law when $\ell\rightarrow \ell_c$ from below ($\gamma_2=1$), while the site and bond nematic susceptibilities diverge only logarithmically ($\gamma_1=\gamma_3=0^+$).

\subsection{Anisotropic case}
\begin{figure}[!t]
\begin{center}
\includegraphics[width=0.6\columnwidth]{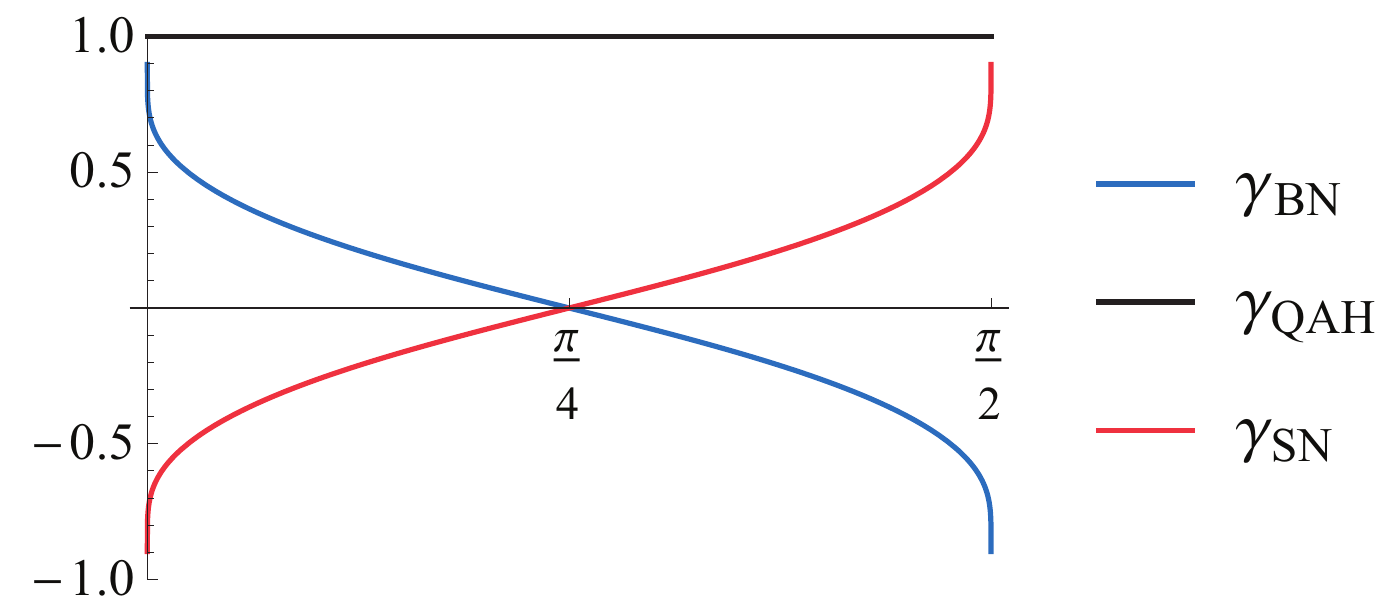}
\end{center}
\caption{Comparison of the susceptibility exponents for the three possible orders in the absence of $C_4$ symmetry. Here ``BN" = bond nematic, ``QAH" = quantum anomalous Hall, and ``SN" = site nematic.}
\label{fig:digrams1}
\end{figure}
The single particle Hamiltonian,
\begin{align}
H_0 = \sum_{|\bk|<{\Lam_0}}\Psi^\dagger_\bk\left(\frac{k^2_x-k^2_y}{2}\tau_3+ \frac{2k_xk_y}{2}\tau_1\right)\Psi_\bk,
\end{align}
is invariant under $\pi/4$ rotations on the $x-y$ plane: $(k_x, k_y ) \mapsto (k'_x + k'_y, k'_x - k'_y)/\sqrt{2}$, and $\Psi_\bk \mapsto \frac{\tau_1 + \tau_3}{\sqrt{2}} \Psi_{\mbf k'}$.
Since the interactions do not possess the symmetry, quantum corrections can in principle remove it by introducing an anisotropy between the two terms.
Here we consider the behavior of the susceptibilities in the presence of such anisotropy,
\begin{eqnarray}
H&=&H_0+H_{int}\\
H_0&=&\sum_{|\bk|<{\Lam_0}}\Psi^\dagger_\bk\left(\cos\eta\frac{k^2_x-k^2_y}{\sqrt{2}}\tau_3+\sin\eta\frac{2k_xk_y}{\sqrt{2}}\tau_1\right)\Psi_\bk\\
H_{int}&=& g\int d^2\br\; \psi_a^\dagger(\br)\psi_b^\dagger(\br)\psi_b(\br)\psi_a(\br)
=\frac{1}{4}g\int d^2\br\; \Psi^\dagger(\br)\tau_2\Psi^*(\br)\Psi^T(\br)\tau_2\Psi(\br)
\end{eqnarray}
where $\eta\in(0,2\pi)$ quantifies the degree of the anisotropy \cite{murray2014}.
Thus, 
\begin{eqnarray}
\int_{-\infty}^{\infty}\frac{d\omega}{2\pi}\int^{\Lam_0}_{\frac{{\Lam_0}}{s}} \frac{d^2\bk}{(2\pi)^2} G_\bk(i\omega) \otimes G_{\mp k}(\mp i\omega) =
\left(\pm a_0(\eta)1\otimes 1+a_1(\eta)\tau_1\otimes \tau_1+a_3(\eta)\tau_3\otimes \tau_3\right)\frac{1}{4\pi}\ln s\nonumber\\
\end{eqnarray}
where
\begin{eqnarray}
a_0(\eta)&=&\frac{\sqrt{2}}{\pi}\frac{K\left(\sqrt{1-\cot^2\eta}\right)}{|\sin\eta|}\\
a_1(\eta)&=&a_3\left(\eta+\frac{\pi}{2}\right)=\frac{\sqrt{2}}{\pi}
\frac{K\left(\sqrt{1-\cot^2\eta}\right)-E\left(\sqrt{1-\cot^2\eta}\right)}{|\sin\eta|(1-\cot^2\eta)}
\end{eqnarray}
and
\begin{eqnarray}
K(x)&=&\int_0^{\frac{\pi}{2}}\frac{d\theta}{\sqrt{1-x^2\sin^2\theta}}\\
E(x)&=&\int_0^{\frac{\pi}{2}} d\theta \sqrt{1-x^2\sin^2\theta}
\end{eqnarray}

Following the same procedure, we find the susceptibility exponents
\begin{eqnarray}
\gamma_2&=&1\\
\gamma_1&=&-\gamma_3=2\frac{a_0(\eta)-a_1(\eta)+a_3(\eta)}{a_0(\eta)+a_1(\eta)+a_3(\eta)}-1\nonumber\\
&=&\frac{1}{\cot^2\eta-1}\left(\frac{1}{\sin^2\eta}-2\frac{E\left(\sqrt{1-\cot^2\eta}\right)}{K\left(\sqrt{1-\cot^2\eta}\right)}\right)
\end{eqnarray}

The susceptibility exponents are plotted as a function of $\eta$ in the Fig.~\ref{fig:digrams1}. 
We deduce that unless the anisotropy is an extreme one, i.e. $\eta = 0$ or $\pi/2$ in which case one of the two terms in $H_0$ is absent, the QAH remains a dominant instability of the QBT semimetal.


\section{PEGP for nematic order} \label{app:pegp-nem}
In this appendix we use the PEGP method to show the absence of a nematic order at weak coupling.
The site-nematic order parameter is
\begin{align}
\hat{\Dl}_{nem} = \sum_{\mbf r} \avg{a_{\mbf r}^{\dag} a_{\mbf r} - b_{\mbf r}^{\dag} b_{\mbf r}} ,
\end{align}
where $a_{\mbf r}$  and $b_{\mbf r}$ are fermion operators, and $\mbf{r}$ labels the unit cell.  
On Fourier transforming we obtain 
\begin{align}
\hat{\Dl}_{nem} = \int \dd{\mbf{k}} \avg{\psi^{\dag}(\bs k) \sig_3 \psi(\bs k)},
\end{align}
where $\psi(\bs k) = \trans{(a(\bs k), b(\bs k))}$, and $\int \dd{\bs k} \equiv \int_{-\pi}^{\pi} \frac{dk_x}{2\pi} \frac{dk_y}{2\pi}$.
Adding $\int \frac{dk_0}{2\pi} J_{nem} \hat{\Dl}_{nem}$ to the action we obtain a $J_{nem}$-dependent propagator,
\begin{align}
G(k; J_{nem}) = \frac{ik_0 + d_1(\bs k) \sig_1 + (J_{nem} + d_3(\bs k))\sig_3 }{k_0^2 + d_1^2(\bs k) + (J_{nem} + d_3(\bs k))^2 }.
\end{align}
Here
\begin{align}
d_1(\bs k) = 4 \cos{\frac{k_x}{2}} \cos{\frac{k_y}{2}}, \qquad
d_3(\bs k) = \cos{k_x} - \cos{k_y}.
\end{align}
The gap,
\begin{align}
\Dl_{nem}(J_{nem}) &\equiv \avg{ \int \dd{k} \psi^{\dag}(k) \sig_3 \psi(k)}
= - \int \dd{k} \tr{\sig_3 G(k;J_{nem})} = - \int \dd{\bs k} \frac{J_{nem} + d_3(\bs k)}{M(\bs k; J_{nem})},
\end{align}
where $\int dk \equiv \int_{-\infty}^{\infty} \frac{dk_0}{2\pi} \int \dd{\bs k}$, and 
\begin{align}
M(\bs k; J_{nem}) = \sqrt{d_1^2(\bs k) + (J_{nem} + d_3(\bs k))^2}.
\end{align}

The total action is
\begin{align}
S[J_{nem}] &= \int \dd{k} \psi^{\dag}(k) ~G^{-1}(k; J_{nem}) ~\psi(k) \nn \\
& + 2 V_2 \int \dd{k_1} \dd{k_2} \dd{q} (\cos{q_x} + \cos{q_y}) \lt[ a^{\dag}(k_1+q) a(k_1) a^{\dag}(k_2-q) a(k_2) + a \rtarw b\rt].
\label{eq:S-unsym}
\end{align}
Upon anti-symmetrizing the interaction vertex we obtain,
\begin{align}
S[J_{nem}] &= \int \dd{k} \psi^{\dag}(k) ~G^{-1}(k; J_{nem}) ~\psi(k) \nn \\
& + 2 V_2 \int \dd{k_1} \dd{k_2} \dd{q} \lt(\sin{\frac{k_{1x} - k_{2x}}{2}} \sin{\frac{k_{1x} - k_{2x} + 2q_x}{2}} + x\rtarw y \rt) \nn \\
& \qquad \times \lt[ a^{\dag}(k_1+q) a(k_1) a^{\dag}(k_2-q) a(k_2) + a \rtarw b\rt].
\label{eq:S-antisym}
\end{align}
Therefore,
\begin{align}
\avg{S_{int}} &= 2 V_2 \int \dd{k_1} \dd{k_2} \dd{q} \lt(\sin{\frac{k_{1x} - k_{2x}}{2}} \sin{\frac{k_{1x} - k_{2x} + 2q_x}{2}} + x\rtarw y \rt) \nn \\
& \quad \times \lt[ \avg{a(k_1) a^{\dag}(k_1+q)} \avg{a(k_2) a^{\dag}(k_2-q)} 
- \avg{a(k_1) a^{\dag}(k_2-q)} \avg{a(k_2) a^{\dag}(k_1+q)} + a \rtarw b\rt].
\end{align}
The first term corresponds to the Hartree diagram, while the last term corresponds to the Fock diagram.
Using the relationships,
\begin{align}
\avg{a(k) a^\dag(k')} = (2\pi)^3 \dl^{(3)}(k - k') ~ G_{11}(k),
\qquad
\avg{b(k) b^\dag(k')} = (2\pi)^3 \dl^{(3)}(k - k') ~ G_{22}(k),
\end{align}
we obtain (using the identity, $\cos{2x} = 1 - 2\sin^2{x} = 2\cos^2{x} - 1$)
\begin{align}
\avg{S_{int}} &= 4 V_2 (2\pi)^3 \dl^{(3)}(0) \int \dd{k_1} \dd{k_2} \lt\{\sin^2\lt({\frac{k_{1x} - k_{2x}}{2}}\rt) + \sin^2\lt({\frac{k_{1y} - k_{2y}}{2}}\rt) \rt\} \nn \\
&\qquad \times \lt[ G_{11}(k_1) G_{11}(k_2) + G_{22}(k_1) G_{22}(k_2)\rt] \\
&= V_2 (2\pi)^3 \dl^{(3)}(0) 
\lt[ 2 \Dl_{nem}^2(J_{nem}) - \lt\{I_x^2(J_{nem}) + I_y^2(J_{nem}) \rt\} \rt],
\label{eq:S-int-1}
\end{align}
where 
\begin{align}
I_x(J_{nem}) &= \int \dd{\bs k} \cos{(k_x)}~ \frac{J_{nem} + d_3(\bs k)}{M(\bs k; J_{nem})}, \\
I_y(J_{nem}) &= \int \dd{\bs k} \cos{(k_y)}~ \frac{J_{nem} + d_3(\bs k)}{M(\bs k; J_{nem})}.
\label{eq:fock}
\end{align}
Therefore, the Gibbs free energy for the interacting theory with only $V_2$ term, up to linear order in $V_2$, is given by
\begin{align}
\mc{G}(\Dl_{nem}) =  \mc{G}_0(\Dl_{nem}) + (2\pi)^3 \dl^{(3)}(0) V_2 \lt[ 2\Dl_{nem}^2(J_{nem}) - I_x^2(J_{nem}) - I_y^2(J_{nem}) \rt],
\end{align}
By exchanging $k_x \ltrtarw k_y$, we note that $I_x(J_{nem}) = - I_y(-J_{nem})$.
Differentiating both sides of \eq{eq:G-nem} with respect to $\Dl_{nem}$ leads to 
\begin{align}
\frac{\mc{G}'(\Dl_{nem})}{(2\pi)^3 \dl^{(3)}(0)} &= - J_{nem} +  2 V_2 \lt[ 2\Dl_{nem}(J_{nem}) \rt. \nn \\
& \qquad \lt. -  J_{nem}'(\Dl_{nem}) \lt\{I_x(J_{nem}) I_x'(J_{nem}) + I_y(J_{nem}) I_y'(J_{nem}) \rt\}\rt].
\label{eq:G-nem'}
\end{align}

The existence of a phase transition at weak coupling is crucially dependent on the presence of a $J_{nem} \ln{J_{nem}}$ term in \eq{eq:G-nem'} arising from $\avg{S_{int}}$.
Here we show that this term is absent due to a cancellation between the Hartree and Fock type diagrams.

We note that \eq{eq:fock} may be written as,
\begin{align}
I_x(J_{nem}) &= \int \dd{\bs k} \lt(2\cos^2{\frac{k_x}{2}} - 1 \rt)~ \frac{J_{nem} + d_3(\bs k)}{M(\bs k; J_{nem})}
= I_{x;1}(J_{nem}) + \Dl_{nem}(J_{nem}),
\label{eq:fock-1}
\end{align}
where
\begin{align}
I_{x;1}(J_{nem})  = 2\int \dd{\bs k} \cos^2{\frac{k_x}{2}} ~ \frac{J_{nem} + d_3(\bs k)}{M(\bs k; J_{nem})}.
\end{align}
Similarly
\begin{align}
I_y(J_{nem}) &= I_{y;1}(J_{nem}) + \Dl_{nem}(J_{nem})
\label{eq:fock-2}
\end{align}
with
\begin{align}
I_{y;1}(J_{nem})  = 2\int \dd{\bs k} \cos^2{\frac{k_y}{2}} ~ \frac{J_{nem} + d_3(\bs k)}{M(\bs k; J_{nem})}.
\end{align}
Therefore, using results in Eqs. \eqref{eq:fock-1} and \eqref{eq:fock-2},
\begin{align}
2 \Dl_{nem}^2 - \{I_x^2 + I_y^2 \} 
= - I_{x;1}^2 - I_{y;1}^2 - 2 \Dl_{nem}(I_{x;1} + I_{y;1}),
\end{align}
where we have suppressed the dependence on $J_{nem}$.
In order to determine the leading order behavior of $I_{n;1}$ in the small $J_{nem}$ limit, we compute $\dow_{J_{nem}}I_{n;1}$,
\begin{align}
\dow_{J_{nem}}I_{x;1}(J_{nem}) = 2\int \dd{\bs k} \cos^2{\frac{k_x}{2}} ~ \frac{d_1^2(\bs k)}{M^{3/2}(\bs k; J_{nem})}.
\end{align}
Therefore,
\begin{align}
\dow_{J_{nem}}I_{x;1}(J_{nem} \rtarw 0) = 2\int \dd{\bs k} \cos^2{\frac{k_x}{2}} ~ \frac{d_1^2(\bs k)}{\lt(2 + \cos{k_x} + \cos{k_y} \rt)^3}.
\end{align}
Near the $\mbf{M}$ point the integrand \begin{align}
\sim \frac{k_x^4 k_y^2 }{(k_x^2+ k_y^2)^3} 
= \cos^4{\theta} ~\sin^2{\theta},
\end{align}
which implies that $\dow_{J_{nem}}I_{x;1}(J_{nem} \rtarw 0)$ is finite, and 
\begin{align}
I_{x;1}(J_{nem}) = I_{y;1}(J_{nem}) = 0.32 J_{nem} + \ordr{J_{nem}^2}.
\end{align}
Thus,
\begin{align}
2 \Dl_{nem}^2 - \{I_x^2 + I_y^2 \} = - 1.28 J_{nem} \Dl_{nem} - \ordr{J_{nem}^2}.
\end{align}
Therefore, $\mc{G}'$ does not vanish for arbitrary (small) $J_{nem}$, which eliminates the presence of a weak coupling instability.


\clearpage

\twocolumngrid
\bibliography{checkerboard}

\end{document}